\documentclass[aps, prd, twocolumn, superscriptaddress, amsfont, graphicx, nofootinbib, preprintnumbers]{revtex4-1}%

\UseRawInputEncoding
\usepackage{color, graphicx, epsfig}
\usepackage{ifpdf}
\usepackage{amsmath}
\usepackage{bm}
\usepackage[english]{babel}
\usepackage{amssymb}
\usepackage{braket}
\usepackage{hyperref}
\usepackage{enumerate}
\usepackage{url}
\usepackage{multirow}
\usepackage{threeparttable}
\usepackage{makecell}
\usepackage{setspace}

\usepackage{slashed}

\usepackage{changes}

\begin{document}

\title{Deep Inelastic Scattering in the Capture of Dark Matter by Neutron Stars}

\author{Liangliang Su}
\email{liangliangsu@njnu.edu.cn}
\affiliation{Department of Physics and Institute of Theoretical Physics, Nanjing Normal University, Nanjing, 210023, China}

\author{Lei Wu}
\email{leiwu@njnu.edu.cn}
\affiliation{Department of Physics and Institute of Theoretical Physics, Nanjing Normal University, Nanjing, 210023, China}

\author{Meiwen Yang}
\email{meiwenyang@njnu.edu.cn}
\affiliation{Department of Physics and Institute of Theoretical Physics, Nanjing Normal University, Nanjing, 210023, China}
\affiliation{Key Laboratory of Dark Matter and Space Astronomy, Purple Mountain Observatory, Chinese Academy of Sciences, Nanjing 210023, China}

\date{\today}

\begin{abstract}

Due to the dense environment, neutron stars (NSs) can serve as an ideal laboratory for studying the interactions between dark matter (DM) and ordinary matter. In the process of DM capture, deep inelastic scattering may dominate over elastic scattering, especially for the DM with a large momentum transfer. In this work, we calculate DM-nucleon deep inelastic scattering via a vector mediator and estimate its contribution to the capture rate. Using the surface temperature of the NSs, we derive the exclusion limits for the DM-nucleon scattering cross section in the mass range, $1~{\rm GeV}<m_{\chi}< 10^{5}~{\rm GeV}$. We find the bounds for DM with the mass $\gtrsim$ 1 GeV can be changed several times after including the deep inelastic scattering contribution. 

\end{abstract}
\pacs{Valid PACS appear here}

\maketitle
\section{Introduction}
A plethora of astronomical observations and cosmological measurements indicate the existence of dark matter in the universe. In recent years, many laboratory experiments have been dedicated to searching for well-theoretically motivated dark matter candidates, such as the Weakly Interacting Massive Particles (WIMPs)~\cite{Lee:1977ua,Jungman:1995df}. However, the null conclusive results about the non-gravitational interaction of DM pose a significant challenge for direct detection~\cite{PandaX-4T:2021bab, LZ:2022ufs, XENON:2023cxc}. 

On the other hand, compact stars serve as unique celestial laboratories, offering an extreme environment to investigate the interactions between dark matter (DM) and Standard Model (SM) particles. As DM traverses through a star, it will scatter with the stellar components, leading to the capture and impacting stellar evolution~\cite{Press:1985ug, Gould:1987ir,Spolyar:2007qv,Fairbairn:2007bn,Freese:2008ur,Taoso:2008kw,Vincent:2015gqa}. The investigations of DM properties through compact astrophysical objects have attracted increasing interest (see e.g. ~\cite{Moskalenko:2007ak,Bertone:2007ae,McCullough:2010ai,Hooper:2010es,deLavallaz:2010wp,Bramante:2013hn,Bell:2013xk,Bertoni:2013bsa,Graham:2015apa,Bramante:2015cua,Amaro-Seoane:2015uny,Graham:2018efk, Cermeno:2018qgu,Janish:2019nkk,Dasgupta:2019juq,Panotopoulos:2020kuo,Dasgupta:2020dik,Horowitz:2020axx,Garani:2020wge,Bose:2022ola,Hardy:2022ufh,Nguyen:2022zwb,Linden:2024uph,Song:2024rru,Yadav:2024xob,Lu:2024kiz,Ema:2024wqr,Liu:2024qbe,Das:2024thc}). 

Notably, owing to the high density of nucleons within neutron stars, the capture rate of DM is significantly increased, which can result in substantial energy loss and make DM become gravitationally bound to the star and accumulate over time. Such processes may affect the evolution of neutron stars; for instance, the accretion of asymmetric DM onto a neutron star could trigger gravitational collapse into a black hole~\cite{Goldman:1989nd, Kouvaris:2010vv, McDermott:2011jp,Kouvaris:2011fi,Bramante:2014zca,Fuller:2014rza, Bramante:2017ulk,Garani:2018kkd,Dasgupta:2020mqg,Giffin:2021kgb,Garani:2021gvc,Bhattacharya:2023stq}. Additionally, DM annihilation can heat neutron stars, maintaining them at higher temperatures~\cite{Kouvaris:2007ay,deLavallaz:2010wp}.

DM kinetic heating of NSs has been proposed to constrain the scattering cross section of DM with ordinary matters~\cite{Baryakhtar:2017dbj, Raj:2017wrv, Camargo:2019wou, Bell:2019pyc,Hamaguchi:2019oev,Garani:2019fpa, Acevedo:2019agu, Joglekar:2019vzy,Joglekar:2020liw}. NSs older than a billion years are sufficiently cold with temperatures $\sim$ 100 K~\cite{Page:2004fy, Yakovlev:2004yr,Baryakhtar:2017dbj}. However, in the presence of DM, the interaction of DM with the NS constituent provides an additional heating mechanism by transferring kinetic energy to the NSs. 
This kinetic heating could potentially raise the temperature of the NSs near Earth to around $\sim 1000 \; \rm K$~\cite{Baryakhtar:2017dbj}. 
The resulting radiation may be detectable by telescopes such as the James Webb Space Telescope~\cite{Gardner:2006ky, Kalirai:2018qfg}, the Thirty Meter Telescope~\cite{TMTInternationalScienceDevelopmentTeamsTMTScienceAdvisoryCommittee:2015pvw}, or the European Extremely Large Telescope~\cite{Baryakhtar:2017dbj,neichel2018overvieweuropeanextremelylarge}. 

Under the strong gravitational pull of the neutron star, the halo DM can be accelerated to velocities close to the speed of light during the capture process. Then, energetic DM will transfer large energy and momentum to neutrons in their scattering process. Therefore, it is essential to consider the internal structure of the neutron and the contribution of inelastic scattering to the capture rate. In previous studies, the calculation of the DM-neutron scattering cross-section primarily focused on elastic scattering. Alternatively, the effects of inelastic scattering were neglected because the elastic scattering was overestimated under the assumption of the first power of the momentum-dependent dipole form factor~\cite{Anzuini:2021lnv}.
In this work, we investigate a simplified fermionic dark matter (DM) model wherein the DM interacts with standard model (SM) particles through the exchange of a dark photon. Concurrently, we will utilize a more accurate electromagnetic form factor of neutron measured by the experiment to calculate the elastic scattering, subsequently estimating the involvement of deep inelastic scattering in DM capture. Our findings reveal that, for DM masses exceeding 1 GeV, the influence of inelastic scattering on the DM capture rate closely approaches and even surpasses that of elastic scattering, particularly in environments characterized by heavier NSs. 

\section{DM-neutron elastic and inelastic scattering}
The halo of dark matter surrounding a neutron star can undergo gravitational acceleration induced by the neutron star, reaching a (semi-)relativistic state. Subsequently， they will deposit most energy via single or multi-scattering, leading to their entrapment within the gravitational confines of the neutron star. This phenomenon is commonly referred to as DM capture by a neutron star. The energy deposition in this process will heat the neutron star changing the surface temperature of a neutron star, which can be observed by some telescopes. In the calculation of DM capture, two significant factors come into play: the equations of state (EoS) characterizing neutron stars and the interactions between DM and the materials constituting the neutron star. For the former, this study employs benchmark neutron star model utilizing the Brussels–Montreal functionals Bsk24~\cite{Pearson:2018tkr,Bell:2020jou}. The accuracy and comprehensiveness of DM-neutron scattering are pivotal, especially concerning the contribution of inelastic scattering involving (semi-)relativistic heavier DM particles.  To facilitate the demonstration of the effects of inelastic scattering, our focus is specifically directed towards DM with masses ranging from $1 \;{\rm GeV}$ to $10^{5}\;{\rm GeV}$,  corresponding to the single scattering regime~\cite{Baryakhtar:2017dbj}.

Firstly, in this section, we will reevaluate the DM-neutron elastic scattering cross-section, incorporating experimentally derived electromagnetic form factors. Furthermore, we will integrate DM-neutron deep inelastic scattering into the calculation of the DM capture rate by NS. We assume that DM couples with neutron via a dark photon, and the corresponding Lagrangian is given by 
\begin{equation}
\mathcal{L}_{int}=g_\chi \bar{\chi} \gamma^\mu \chi A^{\prime}_\mu+ \sum_q g_{q}\bar{q} \gamma^\mu q A^{\prime}_\mu, 
\end{equation}
where $\chi$, $A^{\prime}$ and $q$ denote the DM, dark photon, and quarks, respectively. $g_\chi$ and $g_{q}=\mathcal{Q}_q \epsilon e$ are the coupling constants of DM-dark photon and quarks-dark photon vertex, which $\mathcal{Q}_q$ is the charge number of quarks and $\epsilon$ is the kinetic mixing between dark photon and photon. In general, there is an interaction between the dark photon mediator and the charged leptons. However, the contribution of DM-leptons scattering to the capture rate and then the dark heating is usually secondary~\cite{Joglekar:2019vzy, Joglekar:2020liw}, and will not have a significant impact on our main conclusion. Therefore, in this work, we will only consider the primary contribution of DM capture: DM-neutron scattering.

\subsection{DM-nucleon elastic scattering}
In the process of DM-nucleon scattering, when the energy involved is insufficient to induce nucleon excitation to higher energy states or fragmentation, the resultant interaction is termed DM-nucleon elastic scattering. 
Notably, when the momentum transfer tends toward zero, nucleons can be treated as point-like particles. This would predict the absence of a DM-neutron interaction. As the momentum transfer increases, the internal structure of the nucleon becomes increasingly significant. Thus, the DM-nucleon interaction current is not a simple superposition of DM-quarks interaction current but rather depends on the electromagnetic form factors of nucleon~\cite{Perdrisat:2006hj} like,
\begin{equation}
      \langle \Bar{N}(p_f)|\bar{q} \gamma^\mu q|N(p_i)\rangle=\sum_{q}C_N^q\bar{N}(p_f)\Gamma^{\mu}N(p_i),
 \end{equation}
with 
\begin{equation}
   \Gamma^{\mu} = \gamma^\mu F_1(Q^2)+\frac{i}{2m_N}\sigma^{\mu\nu}q_\nu F_2(Q^2),
\end{equation}
where $p_f$/$p_i$ is the initial/final state nucleon momentum, $Q^2=-q^2$ and $q^2$ is the square of the momentum transfer. The coupling constant between nucleon (neutron or proton) and dark photon is defined as $g_{N}=\sum_q C_N^q g_q=\epsilon e$. 

The functions $F_1(Q^2)$ and $F_2(Q^2)$ represent the Dirac and Pauli form factors respectively, encapsulating the electric and magnetic structure of the nucleon. In previous works, $F_2(Q^2)$ has often been disregarded due to its suppression at $\mathcal{O}(q/m_N)$ for lower momentum transfers, and  $F_1(Q^2)$ is adopted as the dipole form for both neutron and proton. However, in this study, our focus lies on high-energy DM-neutron scattering. Given the potential for substantial momentum transfer inherent in high-energy DM interactions, neglecting the Pauli form factor becomes untenable. Furthermore, the dipole form factor will break down as the momentum transfer approaches zero, as it would erroneously predict non-zero DM-neutron interactions.

Therefore, in this work, we adopt more accurate Dirac and Pauli form factors derived from experimental measurements. These factors are then reformulated in terms of the Sachs electric $G_E$ and magnetic form factors $G_M$ for convenience, that is~\cite{Sachs:1962zzc}, 
\begin{equation}
    \begin{aligned}
          F_1^N(Q^2)&=\frac{G_E^N(Q^2)+ \tau G_M^N(Q^2)}{1+\tau}, \\
        F_2^N(Q^2)&= \frac{G^{N}_M(Q^2)-G_E^N(Q^2)}{1+\tau}, 
    \end{aligned}
\end{equation}
where $\tau \equiv{Q^2}/{4m_n^2}$. The electric form factors and their measurement methods for neutrons and protons are different because of the variations in their charge distributions. Likewise, differences in their magnetic form factors arise from the distributions of their magnetic moments.  For neutrons, the form factors are extracted from electron-deuterium quasi-elastic scattering and here we adopt the form of Ref.~\cite{Kelly:2004hm,Riordan:2010id}
\begin{equation}\label{eq:G_M_E}
\begin{aligned}
     G_M^n(Q^2)&\equiv\mu_n\frac{\sum_{k=0}^1 a_k \tau^k}{1+\sum_{k=1}^3 b_k \tau^k},\\
     G_E^n(Q^2)&=\frac{1.7\tau}{1+3.3\tau}G_D(Q^2), 
\end{aligned}
\end{equation}
where $\mu_n=-1.913$ is the magnetic moment of the neutron. Other parameters are set as $a_0=1$, $a_1=2.33$, $b_1=14.72$, $b_2=24.2$, $b_3=84.1$, and experiments show that $G_D(Q^2)$ can be fitted well with dipole form
\begin{equation}\label{GD}
    G_D(Q^2)=\frac{1}{\left(1+\dfrac{Q^2}{\Lambda^2}\right)^2}, 
\end{equation}
where $\Lambda^2 = 0.71 \;\mathrm{GeV}^2$ is associated with the charge radius. It is noteworthy that the Dirac form factor of neutron $F_{1}^n$ attains a value of zero precisely at $Q^2=0$ for the electromagnetic form factors of Eq.~(\ref{eq:G_M_E}). This characteristic aligns impeccably with the inherent ``neutrality'' of the neutron, but it is not achievable for the dipole form factor. 

With the DM-neutron effective interaction in hands, we can obtain the differential cross section of DM-neutron elastic scattering, which is given by 
\begin{equation}\label{differential elastic scattering cross section}
        \frac{\rm d\sigma}{\rm dcos\theta}=\frac{1}{16\pi}\frac{\alpha(s)}{2 s \alpha(s)-\beta^2(s)}|\mathcal{M}|^2,
\end{equation}
with the amplitude squared
\begin{equation}\label{Elastic scattering amplitude}
    \begin{aligned}
|\mathcal{M}|^2&=\frac{4g_n^2g_\chi^2}{(Q^2+m_{\mathrm{A}^{\prime}}^{2})^2}\{[(s-m_n^2-m_\chi^2)^2\\
&-(s-m_n^2-m_\chi^2)Q^2
     -m_n^2 Q^2]\frac{G_E^{n 2}+\tau G_M^{n 2}}{1+\tau}
     \\&+[\frac{1}{2}Q^4-m_\chi^2Q^2]G_M^{n 2}\},\\
\end{aligned}
\end{equation}
where $\alpha(s)=s-(m_n^2+m_\chi^2)$ and $\beta(s)=\sqrt{\alpha^2(s)-4m_n^2m_\chi^2}$ are the functions of the Mandelstam variables ($s,t, u$), DM mass $m_{\chi}$, and neutron mass $m_n$. The scattering angle $\theta$ in the center of mass frame is related to the momentum transfer $Q^2$.

\begin{figure}[ht]
\includegraphics[width=8cm]{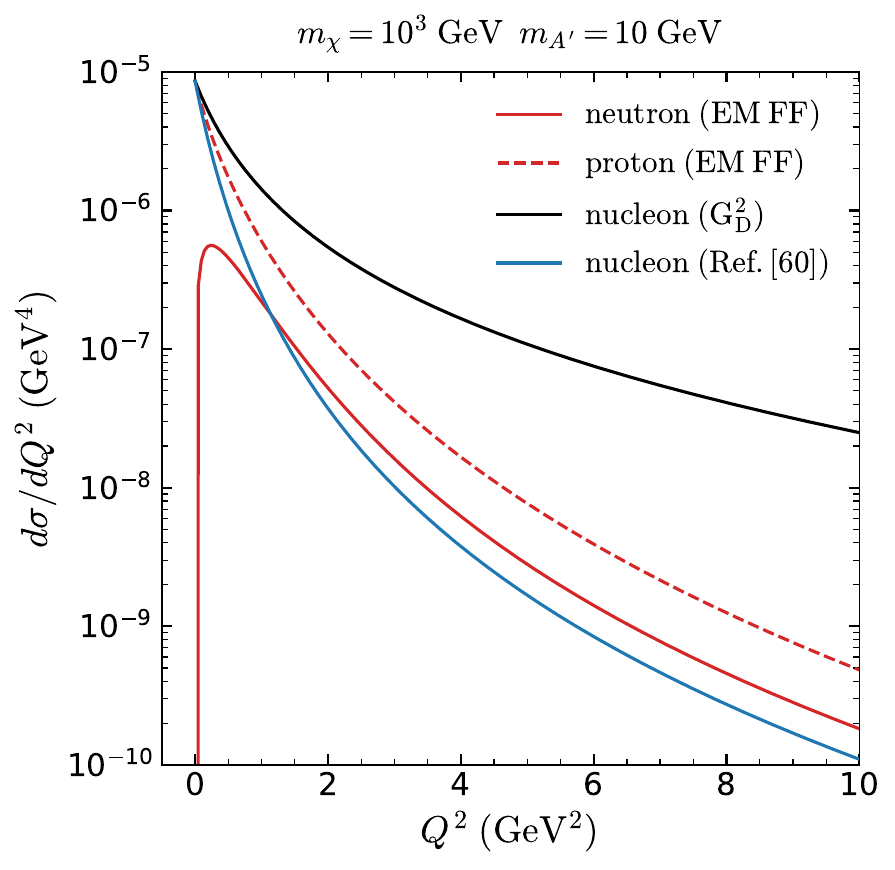}
 \caption{Differential DM-nucleon elastic scattering cross sections in rest frame of neutron as a function of momentum transfer $Q^2$ for various form factors: the electromagnetic form factor(EM FF: red lines), the standard dipole form factor ($|\mathcal{M}|^2 \propto G_D^2$, blue line), and the form of Ref.~\cite{Anzuini:2021lnv} ($|\mathcal{M}|^2 \propto G_D $, black line), respectively. Here, we set the coupling constant $g_N^2 g_\chi^2=1$. }
 \label{fig:form factor}
\end{figure}

To elucidate the impact of the form factor on DM-nucleon scattering cross-sections, we present the differential cross-sections for DM-neutron/proton elastic scattering with various form factors. 
In Fig.~\ref{fig:form factor}, the red solid and dotted lines correspond to DM-neutron and proton scattering utilizing our electromagnetic form factors, respectively.  Additionally, the black solid line depicts the outcomes obtained from employing the first power of the dipole form factor ($|\mathcal{M}|^2 \propto G_D$) as described in Eq.(3.5) of Ref.~\cite{Anzuini:2021lnv}. However, it is imperative to note that the standard squared amplitude should utilize the square of the dipole form factor ($|\mathcal{M}|^2 \propto G_D^2$), represented by the blue solid line. For proton scattering, all employed form factors yield the correct point-like proton-DM differential scattering cross-sections. Conversely, for neutrons, their ``neutrality'' at zero momentum transfer is only reflected in the case of the electromagnetic form factor. 
Within the region of non-zero momentum transfer, the DM-proton scattering cross-sections obtained using the standard dipole form factor are closely five times lower than those derived from the electromagnetic form factor, and there are about two times difference for DM-neutron scattering. This discrepancy arises from the underestimation of the contribution of the magnetic form factor $G_M$ by using the standard dipole form factor. Moreover, the utilization of the form factor presented in Ref.~\cite{Anzuini:2021lnv} tends to overestimate the DM-neutron elastic scattering cross-sections, thereby leading to the neglect of the contribution of inelastic scattering in their work.

\subsection{DM-nucleon Deep Inelastic Scattering}

\begin{figure*}[ht]
\centering
\includegraphics[width=8cm]{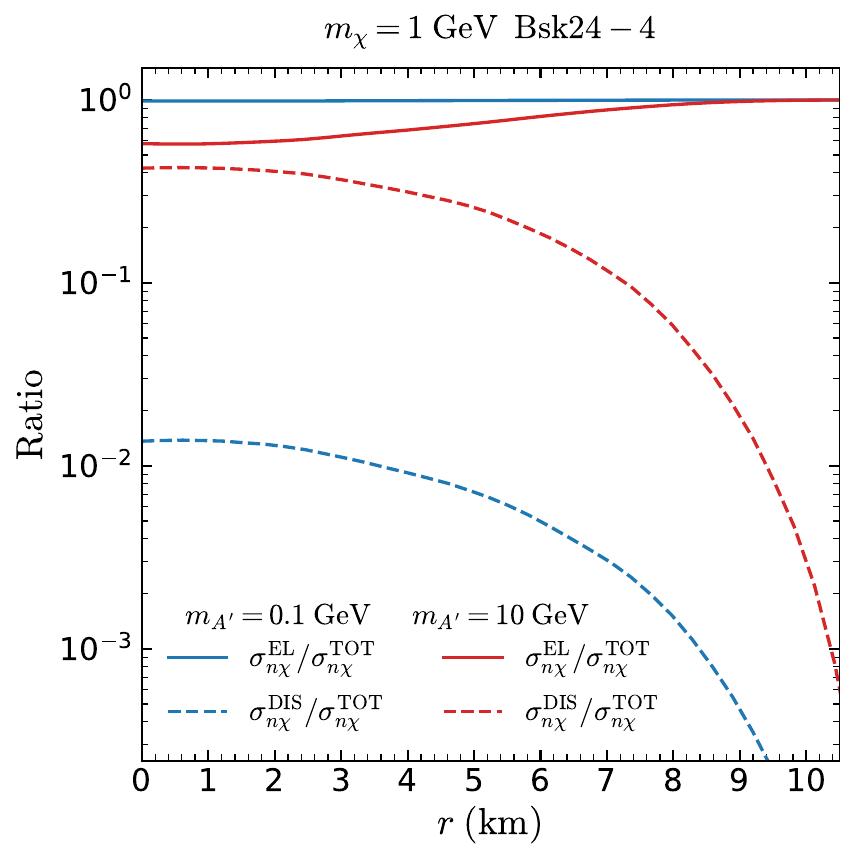}
\includegraphics[width=8cm]{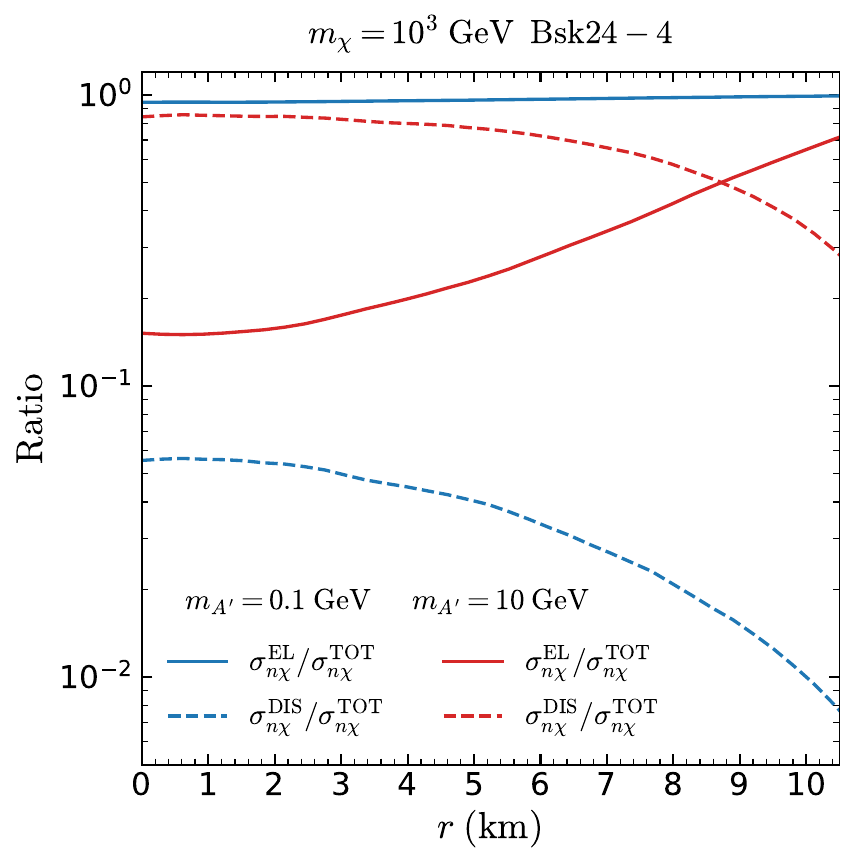}
 \caption{ Ratios of the elastic (EL: solid lines) and deep inelastic scattering (DIS: dashed lines) cross sections to the total cross section (TOT = EL + DIS) as a function of the NS radius, for neutron targets in NS Bsk24-4. The light mediator ($m_{A^\prime} = 0.1$~GeV, blue lines) and heavy mediator ($m_{A^\prime} = 10$~GeV, red lines) are used as benchmark points.}
 \label{fig:Ratio}
\end{figure*}

Indeed, as the transfer momentum of dark matter (DM) scattering escalates ($Q^2\gtrsim 1 \; \rm GeV^2$), the uncertainty principle ($\Delta x \Delta p \gtrsim h$) becomes pertinent. Consequently, high-energy DM particles can function as high-resolution probes, enabling the observation of the internal structure of neutrons. This phenomenon gives rise to DM-neutron deep inelastic scattering, analogous to electron-proton scattering. 

Therefore, we can also use the parton model to calculate the DM-neutron deep inelastic scattering, i.e., the neutron is regarded as the 
aggregation of the spin-1/2 fermion point-like parton (also called quark). Assuming the probability that the $i$-th quark carrying momentum fraction ranges from $x_i$  to  $x_i+dx_i$ is $f_i(x_i)$ that can be abstracted from the nCTEQ15 profile in LHAPDF~\cite{Buckley:2014ana,Kovarik:2015cma},  we can gain the DM-neutron deep inelastic scattering cross section as follows,
\begin{equation}\label{differential inelastic scattering cross section}
   \frac{\mathrm{d}^2\sigma}{\mathrm{d}\cos\theta \mathrm{d}x}=\frac{1}{16\pi}\frac{\alpha(\hat{s})}{2 \hat{s} \alpha(\hat{s})-\beta^2(\hat{s})}\sum_if_i(x_i, Q^2)|\mathcal{M}|^2,
\end{equation}
with the squared amplitude of DM-quark elastic scattering
\begin{equation}\label{inelastic scattering amplitude}
\begin{aligned}
      |\mathcal{M}|^2=\frac{8g_n^2 \mathcal{Q}_i^2 g_{\chi}^2}{(Q^2+m_{A^{\prime}}^{2})^2}( \frac{1}{4}(\hat{s}-Q^2-x_i^2m_n^2-m_\chi^2)^2\\
      +\frac{1}{4}(\hat{s}-x_i^2m_n^2-m_\chi^2)^2-\frac{1}{2}Q^2(x_i^2m_n^2+m_\chi^2)), 
\end{aligned}
\end{equation}
where the Mandelstam variables have been corrected to $\hat{s}=(1-x_i)(m_{\chi}^2-x_i m_n^2)+x_i s$.

The contribution of DM-neutron deep inelastic scattering primarily depends on the momentum transfer from DM to neutrons, a quantity contingent upon the energies of both the DM and neutrons. Considering that the DM mass and the EoS of the neutron star can influence the energies of DM and neutrons at different positions within the neutron star, we examine the ratios of DM-neutron elastic (EL: solid lines) and deep inelastic scattering (DIS: dotted lines) cross sections to the total cross section (TOT=EL+DIS) as functions of neutron star radius for various DM and mediator masses (red lines: $m_{A^{\prime}} =10$~GeV, blue lines: $m_{A^\prime} = 0.1$~GeV), as depicted in Fig.~\ref{fig:Ratio}. It becomes evident that the ratio of inelastic scattering decreases with increasing radius. This phenomenon can be attributed to the neutron energy, which lies within the range $[m_n, m_n+\mu_{F,n}]$, where $\mu_{F,n}$ denotes the chemical potential of neutrons. In our analysis, we select the maximum value of neutron energy as our benchmark point. It is worth noting that $\mu_{F,n}$ decreases with increasing radius, as demonstrated in Fig. 1 of Ref.~\cite{Bell:2020jou}. Consequently, the transfer momentum diminishes as the radius increases, and the significance of inelastic scattering intensifies as the transfer momentum escalates.

Furthermore, we observe that the contribution of DIS for the case of a heavier mediator ($m_{A^{\prime}} = 10$ GeV: red lines) surpasses that of a lighter mediator ($m_{A^{\prime}} = 0.1$ GeV: blue lines) for all DM masses. This phenomenon arises from the momentum-dependent effect of the mediator, i.e., $\mathrm{d} \sigma \propto 1/(Q^2+ m_{A^{\prime}})^2$. For a lighter mediator, the DM-neutron elastic scattering cross-section is enhanced by $1/Q^4$, given that the typical momentum transfer of elastic scattering is smaller than that of deep inelastic scattering. Conversely, for a heavier mediator, this momentum-dependent enhancement diminishes, as $\mathrm{d} \sigma \propto 1/ m_{A^{\prime}}^4$. Consequently, inelastic scattering becomes significant for higher momentum transfers and heavier mediator masses. As a result, within the inner regions of the neutron star, where heavier mediators and DM can transfer larger momentum to neutrons, the dominance of DIS becomes apparent. For example, the ratio of DIS to EL is approximately one order of magnitude larger for heavier mediators, as depicted by the red lines in the right panel of Fig.~\ref{fig:Ratio}.

\section{Capture rate and the heating effect on neutron star}
In this section, we aim to demonstrate how DM-neutron interactions impact DM capture and the associated heating effects. In the Schwarzschild metric around NS, the DM mass rate passing through the NS is given by 
\begin{equation}
    \dot{M}_{\chi}=\pi b^2_{\rm max} u_\chi \rho_{\chi}, 
\end{equation}
with the maximum impact parameter
\begin{equation}\label{bmax}
    {b_{\rm max}}=\frac{R_\star}{u_\chi}\sqrt{\frac{2GM_\star}{R_\star}}(1-\frac{2GM_\star}{R_\star})^{-\frac{1}{2}}=R_\star\frac{v_{esc}}{u_\chi}\gamma_{esc},
\end{equation}
where $\rho_\chi=0.4 \;{\rm GeV/cm^3}$, $G$, $M_{\star}$, and $R_{\star}$ are the DM density around NS, the Newton's constant, mass and radius of NS, respectively. Under the assumption of Maxwell-Boltzmann velocity distribution~\cite{Busoni:2017mhe,Bell:2020jou}, the average relative velocity between NS and DM, denoted as $u_\chi$, can be computed as $330.51 \;\rm km/s$, considering an NS velocity of $v_{\star} = 230 \; \rm km/s$ and a DM velocity dispersion of $v_{d} = 270 \;\rm km/s$. Subsequently, the DM velocity is accelerated to $v_{esc} \approx \sqrt{2 G M_{\star}/R_{\star}}$ at the surface of NS, and the corresponding boost factor is denoted as $\gamma_{esc}$.

The DM surrounding NS will undergo further gravitational attraction, leading to its eventual fall into the NS. During this process, the DM deposits its kinetic energy into the NS through DM-neutron scattering. This mechanism provides an additional heating source for the cooling of NS. The heating rate can be expressed as:
\begin{equation}
\dot{E_k}=\frac{E_s \dot{M_{\chi}}}{m_{\chi}}f, 
\end{equation}
where $E_s=m_{\chi}(\gamma_{esc}-1)$ is the kinetic energy of DM at the surface of NS, which can be almost completely deposited into NS. And $f$ is the fraction of DM capture by NS
\begin{equation}
    f= \min[\frac{C}{C_\star},1], 
\end{equation}
with the geometric limit of the capture rate $C_\star$~\cite{Bell:2018pkk},
\begin{equation}
    C_\star=\frac{\pi R_\star^2(1-B(R_\star))}{v_\star B(R_\star)}\frac{\rho_\chi}{m_\chi}{\rm Erf}(\sqrt{\frac{3}{2}}\frac{v_\star}{v_d}), 
\end{equation}
where $B(r)$ is a function of the radius of the NS, representing the time component of the Schwarzschild metric. This function can be obtained by solving the Tolman-Oppenheimer-Volkoff (TOV) equations governing the structure of NS, and we adopt the results of the Bsk24 neutron star model in this work. Furthermore, we only consider the DM with masses ranging from 1 GeV to $~10^{5}$ GeV. In this DM mass region, the suppression factor of the Pauli blocking effect, $\gamma_{esc} m_{\chi} v_{esc}/p_{F,n}$, is greater than 1 for a typical neutron Fermi momentum $p_{F,n} = 0.4$~GeV~\cite{Baryakhtar:2017dbj}. Additionally, the DM may be captured in a single scattering event. Therefore, the Pauli blocking effect and multiple scattering can be neglected.  Then, the rate of DM capture by NS can be given by~\cite{Bell:2020jou}
\begin{equation}\label{capture rate}
    C=\frac{4\pi}{v_\star}\frac{\rho_\chi}{m_\chi}{\rm Erf}(\sqrt{\frac{3}{2}}\frac{v_\star}{v_d}) \int_0^{R_{\star}} r^2 \frac{\sqrt{1-B(r)}}{B(r)}\Omega^{-}(r){\rm d}r, 
\end{equation}
with the interaction rate $\Omega^{-}(r)$
\begin{equation}
\begin{aligned}
     \Omega^{-}(r)=\frac{1}{4\pi^2}\int \mathrm{d} \cos\theta \mathrm{d} E_n \mathrm{d}s \frac{n_n(r)}{n_{free}(r)}\frac{\mathrm{d}\sigma}{\mathrm{dcos}\theta}\frac{\beta(s)}{\alpha(s)}\frac{E_n}{m_\chi}\\
     \times\sqrt{\frac{B(r)}{1-B(r)}} f_{\rm FD}(E_n, r), 
\end{aligned}
\end{equation}
where $E_n$ is the initial neutron energy. We assume that the neutron in NS obeys the Fermi-Dirac(FD) distribution $f_{\rm FD}(E_n, r)$, which can be approximated as the step function in the zero-temperature approximation for older NSs. In the FD distribution,the realistic number density of neutrons $n_n(r)$ based on the EoS of NS needs to be corrected by the ratio, $\frac{n_n(r)}{n_{free}(r)}$, where $n_{free}(r)=\frac{[\mu_{F,n}(r)(2m_n+\mu_{F,n}(r))]^{3/2}}{3\pi^2}$ in the zero-temperature approximation.

\begin{figure}[ht]
\centering
\includegraphics[width=8cm]{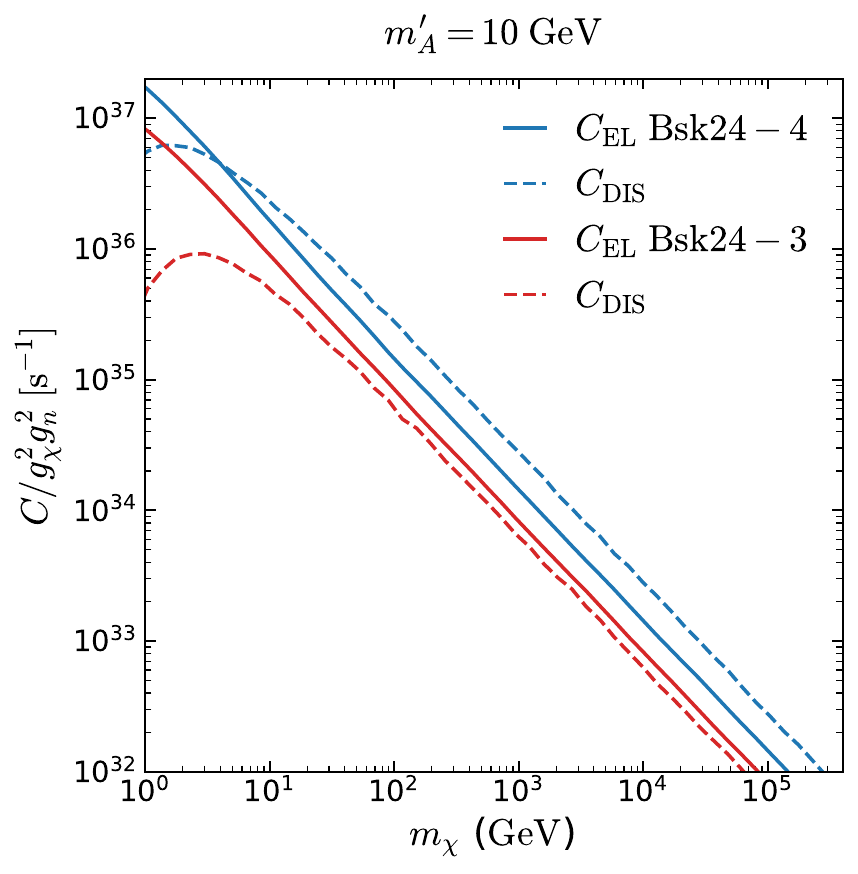}
\caption{The rate of DM capture by DM-neutron elastic scattering (EL: solid lines) and deep inelastic scattering (DIS: dotted lines) in Bsk24-3 (blue lines) and Bsk24-4 (red lines) NS, where we consider a heavier mediator ($m_{A^{\prime}} = 10$ GeV).}
\label{fig:capture_rate}
\end{figure}

The DM-neutron deep inelastic scattering enhances the DM-neutron interaction rate through the differential cross section $\mathrm{d} \sigma/\mathrm{d} \cos \theta$, thereby facilitating a greater capture of DM by NS. Consequently, this leads to more pronounced temperature effects within the NS. For instance, we demonstrate the DM capture rate induced by DM-neutron elastic scattering (EL) and deep inelastic scattering (DIS) via a heavier dark photon ($m_{A^{\prime}} =10$ GeV) for the Bsk24-3 and Bsk24-4 NS model in Fig.~\ref{fig:capture_rate}. In the case of Bsk24-3 NS with a mass of 1.9 $M_\odot$ (red lines), we observe that the contribution of DIS (dotted line) is nearly equivalent to that of EL (solid line) for DM masses exceeding $\sim 10 \;\mathrm{GeV}$. Conversely, for Bsk24-4 NS with a mass of 2.16 $M_\odot$, there is a twofold difference between the contributions of EL and DIS. This discrepancy arises from the ability of both heavier DM and NS to transfer larger momentum to neutrons.

\begin{figure}[h]
\centering
\includegraphics[width=8cm]{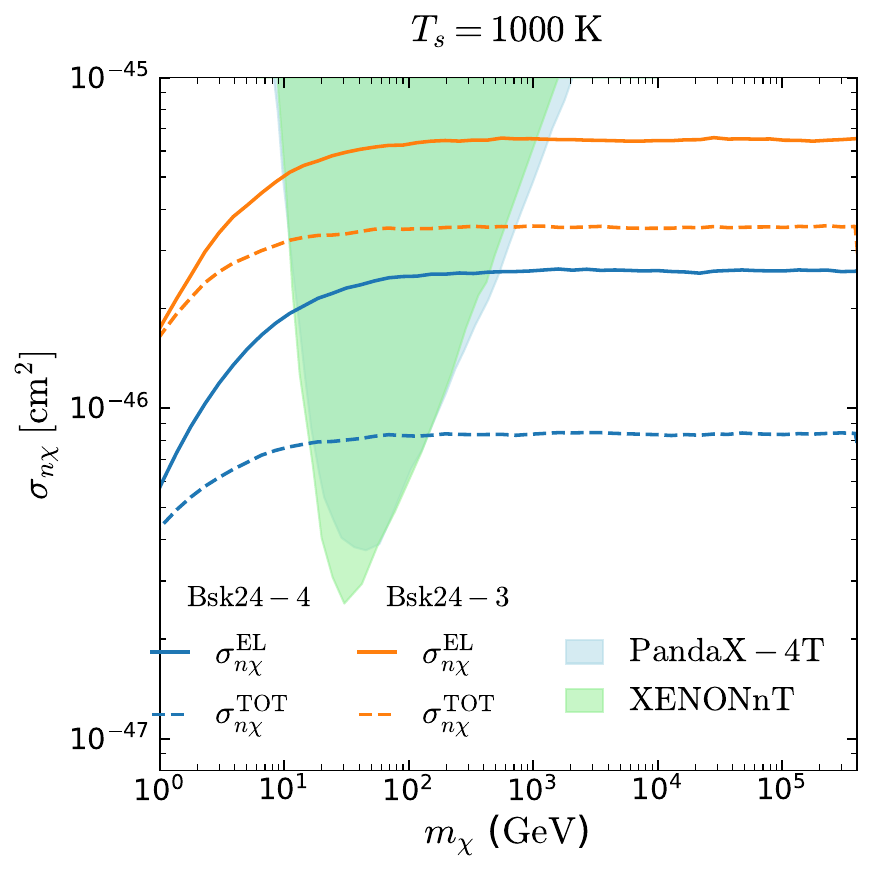}
 \caption{The momentum-independent DM-neutron cross-section $\sigma_{n \chi}$ corresponding to the heating sensitivity of DM capture by NS, as a function of DM mass before (dotted lines) and after (solid lines) including DM-neutron DIS for Bsk24-3 (orange lines) and Bsk24-4 NS (blue lines). Here we set the surface black-body temperature and mediator mass as 1000 K and 10 GeV, respectively. Other bounds from PandaX-4T (shade of light blue)~\cite{PandaX-4T:2021bab}, XENONnT (shade of light green)~\cite{XENON:2023cxc} are shown.}
 \label{sigma_restrict}
\end{figure}

For the young NS, the dark heating makes it difficult to produce an observable signal. However, for old NS, the luminosity of the heating effect from DM capture will eventually balance with the luminosity of black-body radiation (mainly photon emission), which can predict a late-time surface temperature, i.e., 
\begin{equation}
    T_s =\left( \frac{\dot{E}_k}{4 \pi \sigma_{\mathrm{SB}} R_{\star}^2} \right)^{1/4} \sim \mathcal{O}(10^3) f^{1/4} \; \mathrm{K}
\label{eq:Temperature}
\end{equation}
where $\sigma_{\mathrm{SB}}$ is the Stefan-Boltzmann constant. The typical surface temperature of $\mathcal{O}(10^3)$ K provides a promising signal that may be observable by an infrared telescope, particularly for NS with ages on the order of $\mathcal{O}(10)$ Myr, where the black-body temperature is $\lesssim 10^3$ K. Furthermore, some Gyr NSs can cool down to temperatures as low as $\sim 100$ K.
Assuming an NS with a surface black-body temperature of $T_s= 1000$ K is observable, we can determine the corresponding momentum-independent DM-neutron scattering cross section $\sigma_{n \chi} = {g_n^2 g_\chi^2 \mu_n^2}/(\pi m_{\mathrm{A}}^{\prime 4})$ by solving Eq.~\ref{capture rate} and Eq.~\ref{eq:Temperature}, as depicted in Fig.~\ref{sigma_restrict}. Consistent with the conclusions drawn from Fig.~\ref{fig:capture_rate}, we observe that the contribution of DM-neutron DIS in the Bsk24-4 NS (blue lines) exceeds that of the Bsk24-3 NS (orange lines) for DM masses exceeding $ \mathcal{O}(1) \;\mathrm{GeV}$. Particularly, for DM masses ranging from $\mathcal{O}(10)$ GeV to $10^5 \;\mathrm{GeV}$, the DM-neutron scattering cross-section can be constrained to $5\times10^{-46} \rm \; cm^2$ and $10^{-46}\rm \; cm^2$ for 1000 K Bsk24-3 and Bsk24-4, respectively. Additionally, lower NS surface temperatures and heavier neutron stars impose stronger constraints on the DM-neutron scattering cross-section.

\section{Conclusions}
The intricate environments of neutron stars offer an exceptional laboratory for probing dark matter and exploring physics beyond the standard model. For instance, the capture of dark matter by neutron stars presents an effective method to detect dark matter, complementing direct detection experiments. 
During the process of DM capture, halo DM acquires significant energy from the gravitational potential of neutron stars, resulting in substantial energy and momentum transfers in DM-neutron scattering. This characteristic enables DM to serve as a high-resolution probe, capable of probing the internal structure of neutrons. In other words, for high kinetic energy dark matter, the dominant mode of interaction with neutrons is inelastic scattering rather than elastic scattering. Therefore, in this work, we have calculated the DM-neutron elastic scattering cross-section and deep inelastic scattering to investigate the contribution of inelastic scattering to DM capture and heating effects. For DM-neutron elastic scattering, we have employed experimentally derived electromagnetic form factors, which account for the distribution of charge and magnetic moment of neutrons, thereby ensuring neutron neutrality at zero momentum transfer. Finally, findings indicate that the effect of inelastic scattering on DM capture in neutron stars is most pronounced for heavier neutron stars, where the gravitational fields are stronger and thus the maximum transfer momentum is larger. For instance, in heavy neutron stars such as Bsk24-4 NS, considering the effect of inelastic scattering reduces the DM-nucleon scattering cross-section by a factor of three for DM masses ranging from 1 to $10^5 \;\mathrm{GeV}$.

\section{Acknowledgments}
This work is supported by the National Natural Science Foundation of China (NNSFC) No. 12275134, No. 12335005 and No. 12147228.

\appendix

\bibliography{refs}

\begin{thebibliography}{80}%
\makeatletter
\providecommand \@ifxundefined [1]{%
 \@ifx{#1\undefined}
}%
\providecommand \@ifnum [1]{%
 \ifnum #1\expandafter \@firstoftwo
 \else \expandafter \@secondoftwo
 \fi
}%
\providecommand \@ifx [1]{%
 \ifx #1\expandafter \@firstoftwo
 \else \expandafter \@secondoftwo
 \fi
}%
\providecommand \natexlab [1]{#1}%
\providecommand \enquote  [1]{``#1''}%
\providecommand \bibnamefont  [1]{#1}%
\providecommand \bibfnamefont [1]{#1}%
\providecommand \citenamefont [1]{#1}%
\providecommand \href@noop [0]{\@secondoftwo}%
\providecommand \href [0]{\begingroup \@sanitize@url \@href}%
\providecommand \@href[1]{\@@startlink{#1}\@@href}%
\providecommand \@@href[1]{\endgroup#1\@@endlink}%
\providecommand \@sanitize@url [0]{\catcode `\\12\catcode `\$12\catcode
  `\&12\catcode `\#12\catcode `\^12\catcode `\_12\catcode `\%12\relax}%
\providecommand \@@startlink[1]{}%
\providecommand \@@endlink[0]{}%
\providecommand \url  [0]{\begingroup\@sanitize@url \@url }%
\providecommand \@url [1]{\endgroup\@href {#1}{\urlprefix }}%
\providecommand \urlprefix  [0]{URL }%
\providecommand \Eprint [0]{\href }%
\providecommand \doibase [0]{http://dx.doi.org/}%
\providecommand \selectlanguage [0]{\@gobble}%
\providecommand \bibinfo  [0]{\@secondoftwo}%
\providecommand \bibfield  [0]{\@secondoftwo}%
\providecommand \translation [1]{[#1]}%
\providecommand \BibitemOpen [0]{}%
\providecommand \bibitemStop [0]{}%
\providecommand \bibitemNoStop [0]{.\EOS\space}%
\providecommand \EOS [0]{\spacefactor3000\relax}%
\providecommand \BibitemShut  [1]{\csname bibitem#1\endcsname}%
\let\auto@bib@innerbib\@empty
\bibitem [{\citenamefont {Lee}\ and\ \citenamefont
  {Weinberg}(1977)}]{Lee:1977ua}%
  \BibitemOpen
  \bibfield  {author} {\bibinfo {author} {\bibfnamefont {B.~W.}\ \bibnamefont
  {Lee}}\ and\ \bibinfo {author} {\bibfnamefont {S.}~\bibnamefont {Weinberg}},\
  }\href {\doibase 10.1103/PhysRevLett.39.165} {\bibfield  {journal} {\bibinfo
  {journal} {Phys. Rev. Lett.}\ }\textbf {\bibinfo {volume} {39}},\ \bibinfo
  {pages} {165} (\bibinfo {year} {1977})}\BibitemShut {NoStop}%
\bibitem [{\citenamefont {Jungman}\ \emph {et~al.}(1996)\citenamefont
  {Jungman}, \citenamefont {Kamionkowski},\ and\ \citenamefont
  {Griest}}]{Jungman:1995df}%
  \BibitemOpen
  \bibfield  {author} {\bibinfo {author} {\bibfnamefont {G.}~\bibnamefont
  {Jungman}}, \bibinfo {author} {\bibfnamefont {M.}~\bibnamefont
  {Kamionkowski}}, \ and\ \bibinfo {author} {\bibfnamefont {K.}~\bibnamefont
  {Griest}},\ }\href {\doibase 10.1016/0370-1573(95)00058-5} {\bibfield
  {journal} {\bibinfo  {journal} {Phys. Rept.}\ }\textbf {\bibinfo {volume}
  {267}},\ \bibinfo {pages} {195} (\bibinfo {year} {1996})},\ \Eprint
  {http://arxiv.org/abs/hep-ph/9506380} {arXiv:hep-ph/9506380} \BibitemShut
  {NoStop}%
\bibitem [{\citenamefont {Meng}\ \emph {et~al.}(2021)\citenamefont {Meng} \emph
  {et~al.}}]{PandaX-4T:2021bab}%
  \BibitemOpen
  \bibfield  {author} {\bibinfo {author} {\bibfnamefont {Y.}~\bibnamefont
  {Meng}} \emph {et~al.} (\bibinfo {collaboration} {PandaX-4T}),\ }\href
  {\doibase 10.1103/PhysRevLett.127.261802} {\bibfield  {journal} {\bibinfo
  {journal} {Phys. Rev. Lett.}\ }\textbf {\bibinfo {volume} {127}},\ \bibinfo
  {pages} {261802} (\bibinfo {year} {2021})},\ \Eprint
  {http://arxiv.org/abs/2107.13438} {arXiv:2107.13438 [hep-ex]} \BibitemShut
  {NoStop}%
\bibitem [{\citenamefont {Aalbers}\ \emph {et~al.}(2022)\citenamefont {Aalbers}
  \emph {et~al.}}]{LZ:2022ufs}%
  \BibitemOpen
  \bibfield  {author} {\bibinfo {author} {\bibfnamefont {J.}~\bibnamefont
  {Aalbers}} \emph {et~al.} (\bibinfo {collaboration} {LZ}),\ }\href@noop {} {\
   (\bibinfo {year} {2022})},\ \Eprint {http://arxiv.org/abs/2207.03764}
  {arXiv:2207.03764 [hep-ex]} \BibitemShut {NoStop}%
\bibitem [{\citenamefont {Aprile}\ \emph {et~al.}(2023)\citenamefont {Aprile}
  \emph {et~al.}}]{XENON:2023cxc}%
  \BibitemOpen
  \bibfield  {author} {\bibinfo {author} {\bibfnamefont {E.}~\bibnamefont
  {Aprile}} \emph {et~al.} (\bibinfo {collaboration} {XENON}),\ }\href
  {\doibase 10.1103/PhysRevLett.131.041003} {\bibfield  {journal} {\bibinfo
  {journal} {Phys. Rev. Lett.}\ }\textbf {\bibinfo {volume} {131}},\ \bibinfo
  {pages} {041003} (\bibinfo {year} {2023})},\ \Eprint
  {http://arxiv.org/abs/2303.14729} {arXiv:2303.14729 [hep-ex]} \BibitemShut
  {NoStop}%
\bibitem [{\citenamefont {Press}\ and\ \citenamefont
  {Spergel}(1985)}]{Press:1985ug}%
  \BibitemOpen
  \bibfield  {author} {\bibinfo {author} {\bibfnamefont {W.~H.}\ \bibnamefont
  {Press}}\ and\ \bibinfo {author} {\bibfnamefont {D.~N.}\ \bibnamefont
  {Spergel}},\ }\href {\doibase 10.1086/163485} {\bibfield  {journal} {\bibinfo
   {journal} {Astrophys. J.}\ }\textbf {\bibinfo {volume} {296}},\ \bibinfo
  {pages} {679} (\bibinfo {year} {1985})}\BibitemShut {NoStop}%
\bibitem [{\citenamefont {Gould}(1987)}]{Gould:1987ir}%
  \BibitemOpen
  \bibfield  {author} {\bibinfo {author} {\bibfnamefont {A.}~\bibnamefont
  {Gould}},\ }\href {\doibase 10.1086/165653} {\bibfield  {journal} {\bibinfo
  {journal} {Astrophys. J.}\ }\textbf {\bibinfo {volume} {321}},\ \bibinfo
  {pages} {571} (\bibinfo {year} {1987})}\BibitemShut {NoStop}%
\bibitem [{\citenamefont {Spolyar}\ \emph {et~al.}(2008)\citenamefont
  {Spolyar}, \citenamefont {Freese},\ and\ \citenamefont
  {Gondolo}}]{Spolyar:2007qv}%
  \BibitemOpen
  \bibfield  {author} {\bibinfo {author} {\bibfnamefont {D.}~\bibnamefont
  {Spolyar}}, \bibinfo {author} {\bibfnamefont {K.}~\bibnamefont {Freese}}, \
  and\ \bibinfo {author} {\bibfnamefont {P.}~\bibnamefont {Gondolo}},\ }\href
  {\doibase 10.1103/PhysRevLett.100.051101} {\bibfield  {journal} {\bibinfo
  {journal} {Phys. Rev. Lett.}\ }\textbf {\bibinfo {volume} {100}},\ \bibinfo
  {pages} {051101} (\bibinfo {year} {2008})},\ \Eprint
  {http://arxiv.org/abs/0705.0521} {arXiv:0705.0521 [astro-ph]} \BibitemShut
  {NoStop}%
\bibitem [{\citenamefont {Fairbairn}\ \emph {et~al.}(2008)\citenamefont
  {Fairbairn}, \citenamefont {Scott},\ and\ \citenamefont
  {Edsjo}}]{Fairbairn:2007bn}%
  \BibitemOpen
  \bibfield  {author} {\bibinfo {author} {\bibfnamefont {M.}~\bibnamefont
  {Fairbairn}}, \bibinfo {author} {\bibfnamefont {P.}~\bibnamefont {Scott}}, \
  and\ \bibinfo {author} {\bibfnamefont {J.}~\bibnamefont {Edsjo}},\ }\href
  {\doibase 10.1103/PhysRevD.77.047301} {\bibfield  {journal} {\bibinfo
  {journal} {Phys. Rev. D}\ }\textbf {\bibinfo {volume} {77}},\ \bibinfo
  {pages} {047301} (\bibinfo {year} {2008})},\ \Eprint
  {http://arxiv.org/abs/0710.3396} {arXiv:0710.3396 [astro-ph]} \BibitemShut
  {NoStop}%
\bibitem [{\citenamefont {Freese}\ \emph {et~al.}(2008)\citenamefont {Freese},
  \citenamefont {Spolyar},\ and\ \citenamefont {Aguirre}}]{Freese:2008ur}%
  \BibitemOpen
  \bibfield  {author} {\bibinfo {author} {\bibfnamefont {K.}~\bibnamefont
  {Freese}}, \bibinfo {author} {\bibfnamefont {D.}~\bibnamefont {Spolyar}}, \
  and\ \bibinfo {author} {\bibfnamefont {A.}~\bibnamefont {Aguirre}},\ }\href
  {\doibase 10.1088/1475-7516/2008/11/014} {\bibfield  {journal} {\bibinfo
  {journal} {JCAP}\ }\textbf {\bibinfo {volume} {11}},\ \bibinfo {pages} {014}
  (\bibinfo {year} {2008})},\ \Eprint {http://arxiv.org/abs/0802.1724}
  {arXiv:0802.1724 [astro-ph]} \BibitemShut {NoStop}%
\bibitem [{\citenamefont {Taoso}\ \emph {et~al.}(2008)\citenamefont {Taoso},
  \citenamefont {Bertone}, \citenamefont {Meynet},\ and\ \citenamefont
  {Ekstrom}}]{Taoso:2008kw}%
  \BibitemOpen
  \bibfield  {author} {\bibinfo {author} {\bibfnamefont {M.}~\bibnamefont
  {Taoso}}, \bibinfo {author} {\bibfnamefont {G.}~\bibnamefont {Bertone}},
  \bibinfo {author} {\bibfnamefont {G.}~\bibnamefont {Meynet}}, \ and\ \bibinfo
  {author} {\bibfnamefont {S.}~\bibnamefont {Ekstrom}},\ }\href {\doibase
  10.1103/PhysRevD.78.123510} {\bibfield  {journal} {\bibinfo  {journal} {Phys.
  Rev. D}\ }\textbf {\bibinfo {volume} {78}},\ \bibinfo {pages} {123510}
  (\bibinfo {year} {2008})},\ \Eprint {http://arxiv.org/abs/0806.2681}
  {arXiv:0806.2681 [astro-ph]} \BibitemShut {NoStop}%
\bibitem [{\citenamefont {Vincent}\ \emph {et~al.}(2015)\citenamefont
  {Vincent}, \citenamefont {Serenelli},\ and\ \citenamefont
  {Scott}}]{Vincent:2015gqa}%
  \BibitemOpen
  \bibfield  {author} {\bibinfo {author} {\bibfnamefont {A.~C.}\ \bibnamefont
  {Vincent}}, \bibinfo {author} {\bibfnamefont {A.}~\bibnamefont {Serenelli}},
  \ and\ \bibinfo {author} {\bibfnamefont {P.}~\bibnamefont {Scott}},\ }\href
  {\doibase 10.1088/1475-7516/2015/08/040} {\bibfield  {journal} {\bibinfo
  {journal} {JCAP}\ }\textbf {\bibinfo {volume} {08}},\ \bibinfo {pages} {040}
  (\bibinfo {year} {2015})},\ \Eprint {http://arxiv.org/abs/1504.04378}
  {arXiv:1504.04378 [hep-ph]} \BibitemShut {NoStop}%
\bibitem [{\citenamefont {Moskalenko}\ and\ \citenamefont
  {Wai}(2007)}]{Moskalenko:2007ak}%
  \BibitemOpen
  \bibfield  {author} {\bibinfo {author} {\bibfnamefont {I.~V.}\ \bibnamefont
  {Moskalenko}}\ and\ \bibinfo {author} {\bibfnamefont {L.~L.}\ \bibnamefont
  {Wai}},\ }\href {\doibase 10.1086/516708} {\bibfield  {journal} {\bibinfo
  {journal} {Astrophys. J. Lett.}\ }\textbf {\bibinfo {volume} {659}},\
  \bibinfo {pages} {L29} (\bibinfo {year} {2007})},\ \Eprint
  {http://arxiv.org/abs/astro-ph/0702654} {arXiv:astro-ph/0702654} \BibitemShut
  {NoStop}%
\bibitem [{\citenamefont {Bertone}\ and\ \citenamefont
  {Fairbairn}(2008)}]{Bertone:2007ae}%
  \BibitemOpen
  \bibfield  {author} {\bibinfo {author} {\bibfnamefont {G.}~\bibnamefont
  {Bertone}}\ and\ \bibinfo {author} {\bibfnamefont {M.}~\bibnamefont
  {Fairbairn}},\ }\href {\doibase 10.1103/PhysRevD.77.043515} {\bibfield
  {journal} {\bibinfo  {journal} {Phys. Rev. D}\ }\textbf {\bibinfo {volume}
  {77}},\ \bibinfo {pages} {043515} (\bibinfo {year} {2008})},\ \Eprint
  {http://arxiv.org/abs/0709.1485} {arXiv:0709.1485 [astro-ph]} \BibitemShut
  {NoStop}%
\bibitem [{\citenamefont {McCullough}\ and\ \citenamefont
  {Fairbairn}(2010)}]{McCullough:2010ai}%
  \BibitemOpen
  \bibfield  {author} {\bibinfo {author} {\bibfnamefont {M.}~\bibnamefont
  {McCullough}}\ and\ \bibinfo {author} {\bibfnamefont {M.}~\bibnamefont
  {Fairbairn}},\ }\href {\doibase 10.1103/PhysRevD.81.083520} {\bibfield
  {journal} {\bibinfo  {journal} {Phys. Rev. D}\ }\textbf {\bibinfo {volume}
  {81}},\ \bibinfo {pages} {083520} (\bibinfo {year} {2010})},\ \Eprint
  {http://arxiv.org/abs/1001.2737} {arXiv:1001.2737 [hep-ph]} \BibitemShut
  {NoStop}%
\bibitem [{\citenamefont {Hooper}\ \emph {et~al.}(2010)\citenamefont {Hooper},
  \citenamefont {Spolyar}, \citenamefont {Vallinotto},\ and\ \citenamefont
  {Gnedin}}]{Hooper:2010es}%
  \BibitemOpen
  \bibfield  {author} {\bibinfo {author} {\bibfnamefont {D.}~\bibnamefont
  {Hooper}}, \bibinfo {author} {\bibfnamefont {D.}~\bibnamefont {Spolyar}},
  \bibinfo {author} {\bibfnamefont {A.}~\bibnamefont {Vallinotto}}, \ and\
  \bibinfo {author} {\bibfnamefont {N.~Y.}\ \bibnamefont {Gnedin}},\ }\href
  {\doibase 10.1103/PhysRevD.81.103531} {\bibfield  {journal} {\bibinfo
  {journal} {Phys. Rev. D}\ }\textbf {\bibinfo {volume} {81}},\ \bibinfo
  {pages} {103531} (\bibinfo {year} {2010})},\ \Eprint
  {http://arxiv.org/abs/1002.0005} {arXiv:1002.0005 [hep-ph]} \BibitemShut
  {NoStop}%
\bibitem [{\citenamefont {de~Lavallaz}\ and\ \citenamefont
  {Fairbairn}(2010)}]{deLavallaz:2010wp}%
  \BibitemOpen
  \bibfield  {author} {\bibinfo {author} {\bibfnamefont {A.}~\bibnamefont
  {de~Lavallaz}}\ and\ \bibinfo {author} {\bibfnamefont {M.}~\bibnamefont
  {Fairbairn}},\ }\href {\doibase 10.1103/PhysRevD.81.123521} {\bibfield
  {journal} {\bibinfo  {journal} {Phys. Rev. D}\ }\textbf {\bibinfo {volume}
  {81}},\ \bibinfo {pages} {123521} (\bibinfo {year} {2010})},\ \Eprint
  {http://arxiv.org/abs/1004.0629} {arXiv:1004.0629 [astro-ph.GA]} \BibitemShut
  {NoStop}%
\bibitem [{\citenamefont {Bramante}\ \emph {et~al.}(2013)\citenamefont
  {Bramante}, \citenamefont {Fukushima},\ and\ \citenamefont
  {Kumar}}]{Bramante:2013hn}%
  \BibitemOpen
  \bibfield  {author} {\bibinfo {author} {\bibfnamefont {J.}~\bibnamefont
  {Bramante}}, \bibinfo {author} {\bibfnamefont {K.}~\bibnamefont {Fukushima}},
  \ and\ \bibinfo {author} {\bibfnamefont {J.}~\bibnamefont {Kumar}},\ }\href
  {\doibase 10.1103/PhysRevD.87.055012} {\bibfield  {journal} {\bibinfo
  {journal} {Phys. Rev. D}\ }\textbf {\bibinfo {volume} {87}},\ \bibinfo
  {pages} {055012} (\bibinfo {year} {2013})},\ \Eprint
  {http://arxiv.org/abs/1301.0036} {arXiv:1301.0036 [hep-ph]} \BibitemShut
  {NoStop}%
\bibitem [{\citenamefont {Bell}\ \emph {et~al.}(2013)\citenamefont {Bell},
  \citenamefont {Melatos},\ and\ \citenamefont {Petraki}}]{Bell:2013xk}%
  \BibitemOpen
  \bibfield  {author} {\bibinfo {author} {\bibfnamefont {N.~F.}\ \bibnamefont
  {Bell}}, \bibinfo {author} {\bibfnamefont {A.}~\bibnamefont {Melatos}}, \
  and\ \bibinfo {author} {\bibfnamefont {K.}~\bibnamefont {Petraki}},\ }\href
  {\doibase 10.1103/PhysRevD.87.123507} {\bibfield  {journal} {\bibinfo
  {journal} {Phys. Rev. D}\ }\textbf {\bibinfo {volume} {87}},\ \bibinfo
  {pages} {123507} (\bibinfo {year} {2013})},\ \Eprint
  {http://arxiv.org/abs/1301.6811} {arXiv:1301.6811 [hep-ph]} \BibitemShut
  {NoStop}%
\bibitem [{\citenamefont {Bertoni}\ \emph {et~al.}(2013)\citenamefont
  {Bertoni}, \citenamefont {Nelson},\ and\ \citenamefont
  {Reddy}}]{Bertoni:2013bsa}%
  \BibitemOpen
  \bibfield  {author} {\bibinfo {author} {\bibfnamefont {B.}~\bibnamefont
  {Bertoni}}, \bibinfo {author} {\bibfnamefont {A.~E.}\ \bibnamefont {Nelson}},
  \ and\ \bibinfo {author} {\bibfnamefont {S.}~\bibnamefont {Reddy}},\ }\href
  {\doibase 10.1103/PhysRevD.88.123505} {\bibfield  {journal} {\bibinfo
  {journal} {Phys. Rev. D}\ }\textbf {\bibinfo {volume} {88}},\ \bibinfo
  {pages} {123505} (\bibinfo {year} {2013})},\ \Eprint
  {http://arxiv.org/abs/1309.1721} {arXiv:1309.1721 [hep-ph]} \BibitemShut
  {NoStop}%
\bibitem [{\citenamefont {Graham}\ \emph {et~al.}(2015)\citenamefont {Graham},
  \citenamefont {Rajendran},\ and\ \citenamefont {Varela}}]{Graham:2015apa}%
  \BibitemOpen
  \bibfield  {author} {\bibinfo {author} {\bibfnamefont {P.~W.}\ \bibnamefont
  {Graham}}, \bibinfo {author} {\bibfnamefont {S.}~\bibnamefont {Rajendran}}, \
  and\ \bibinfo {author} {\bibfnamefont {J.}~\bibnamefont {Varela}},\ }\href
  {\doibase 10.1103/PhysRevD.92.063007} {\bibfield  {journal} {\bibinfo
  {journal} {Phys. Rev. D}\ }\textbf {\bibinfo {volume} {92}},\ \bibinfo
  {pages} {063007} (\bibinfo {year} {2015})},\ \Eprint
  {http://arxiv.org/abs/1505.04444} {arXiv:1505.04444 [hep-ph]} \BibitemShut
  {NoStop}%
\bibitem [{\citenamefont {Bramante}(2015)}]{Bramante:2015cua}%
  \BibitemOpen
  \bibfield  {author} {\bibinfo {author} {\bibfnamefont {J.}~\bibnamefont
  {Bramante}},\ }\href {\doibase 10.1103/PhysRevLett.115.141301} {\bibfield
  {journal} {\bibinfo  {journal} {Phys. Rev. Lett.}\ }\textbf {\bibinfo
  {volume} {115}},\ \bibinfo {pages} {141301} (\bibinfo {year} {2015})},\
  \Eprint {http://arxiv.org/abs/1505.07464} {arXiv:1505.07464 [hep-ph]}
  \BibitemShut {NoStop}%
\bibitem [{\citenamefont {Amaro-Seoane}\ \emph {et~al.}(2016)\citenamefont
  {Amaro-Seoane}, \citenamefont {Casanellas}, \citenamefont {Sch\"odel},
  \citenamefont {Davidson},\ and\ \citenamefont
  {Cuadra}}]{Amaro-Seoane:2015uny}%
  \BibitemOpen
  \bibfield  {author} {\bibinfo {author} {\bibfnamefont {P.}~\bibnamefont
  {Amaro-Seoane}}, \bibinfo {author} {\bibfnamefont {J.}~\bibnamefont
  {Casanellas}}, \bibinfo {author} {\bibfnamefont {R.}~\bibnamefont
  {Sch\"odel}}, \bibinfo {author} {\bibfnamefont {E.}~\bibnamefont {Davidson}},
  \ and\ \bibinfo {author} {\bibfnamefont {J.}~\bibnamefont {Cuadra}},\ }\href
  {\doibase 10.1093/mnras/stw433} {\bibfield  {journal} {\bibinfo  {journal}
  {Mon. Not. Roy. Astron. Soc.}\ }\textbf {\bibinfo {volume} {459}},\ \bibinfo
  {pages} {695} (\bibinfo {year} {2016})},\ \Eprint
  {http://arxiv.org/abs/1512.00456} {arXiv:1512.00456 [astro-ph.CO]}
  \BibitemShut {NoStop}%
\bibitem [{\citenamefont {Graham}\ \emph {et~al.}(2018)\citenamefont {Graham},
  \citenamefont {Janish}, \citenamefont {Narayan}, \citenamefont {Rajendran},\
  and\ \citenamefont {Riggins}}]{Graham:2018efk}%
  \BibitemOpen
  \bibfield  {author} {\bibinfo {author} {\bibfnamefont {P.~W.}\ \bibnamefont
  {Graham}}, \bibinfo {author} {\bibfnamefont {R.}~\bibnamefont {Janish}},
  \bibinfo {author} {\bibfnamefont {V.}~\bibnamefont {Narayan}}, \bibinfo
  {author} {\bibfnamefont {S.}~\bibnamefont {Rajendran}}, \ and\ \bibinfo
  {author} {\bibfnamefont {P.}~\bibnamefont {Riggins}},\ }\href {\doibase
  10.1103/PhysRevD.98.115027} {\bibfield  {journal} {\bibinfo  {journal} {Phys.
  Rev. D}\ }\textbf {\bibinfo {volume} {98}},\ \bibinfo {pages} {115027}
  (\bibinfo {year} {2018})},\ \Eprint {http://arxiv.org/abs/1805.07381}
  {arXiv:1805.07381 [hep-ph]} \BibitemShut {NoStop}%
\bibitem [{\citenamefont {Cerme\~no}\ and\ \citenamefont
  {P\'erez-Garc\'\i{}a}(2018)}]{Cermeno:2018qgu}%
  \BibitemOpen
  \bibfield  {author} {\bibinfo {author} {\bibfnamefont {M.}~\bibnamefont
  {Cerme\~no}}\ and\ \bibinfo {author} {\bibfnamefont {M.~A.}\ \bibnamefont
  {P\'erez-Garc\'\i{}a}},\ }\href {\doibase 10.1103/PhysRevD.98.063002}
  {\bibfield  {journal} {\bibinfo  {journal} {Phys. Rev. D}\ }\textbf {\bibinfo
  {volume} {98}},\ \bibinfo {pages} {063002} (\bibinfo {year} {2018})},\
  \Eprint {http://arxiv.org/abs/1807.03318} {arXiv:1807.03318 [hep-ph]}
  \BibitemShut {NoStop}%
\bibitem [{\citenamefont {Janish}\ \emph {et~al.}(2019)\citenamefont {Janish},
  \citenamefont {Narayan},\ and\ \citenamefont {Riggins}}]{Janish:2019nkk}%
  \BibitemOpen
  \bibfield  {author} {\bibinfo {author} {\bibfnamefont {R.}~\bibnamefont
  {Janish}}, \bibinfo {author} {\bibfnamefont {V.}~\bibnamefont {Narayan}}, \
  and\ \bibinfo {author} {\bibfnamefont {P.}~\bibnamefont {Riggins}},\ }\href
  {\doibase 10.1103/PhysRevD.100.035008} {\bibfield  {journal} {\bibinfo
  {journal} {Phys. Rev. D}\ }\textbf {\bibinfo {volume} {100}},\ \bibinfo
  {pages} {035008} (\bibinfo {year} {2019})},\ \Eprint
  {http://arxiv.org/abs/1905.00395} {arXiv:1905.00395 [hep-ph]} \BibitemShut
  {NoStop}%
\bibitem [{\citenamefont {Dasgupta}\ \emph {et~al.}(2019)\citenamefont
  {Dasgupta}, \citenamefont {Gupta},\ and\ \citenamefont
  {Ray}}]{Dasgupta:2019juq}%
  \BibitemOpen
  \bibfield  {author} {\bibinfo {author} {\bibfnamefont {B.}~\bibnamefont
  {Dasgupta}}, \bibinfo {author} {\bibfnamefont {A.}~\bibnamefont {Gupta}}, \
  and\ \bibinfo {author} {\bibfnamefont {A.}~\bibnamefont {Ray}},\ }\href
  {\doibase 10.1088/1475-7516/2019/08/018} {\bibfield  {journal} {\bibinfo
  {journal} {JCAP}\ }\textbf {\bibinfo {volume} {08}},\ \bibinfo {pages} {018}
  (\bibinfo {year} {2019})},\ \Eprint {http://arxiv.org/abs/1906.04204}
  {arXiv:1906.04204 [hep-ph]} \BibitemShut {NoStop}%
\bibitem [{\citenamefont {Panotopoulos}\ and\ \citenamefont
  {Lopes}(2020)}]{Panotopoulos:2020kuo}%
  \BibitemOpen
  \bibfield  {author} {\bibinfo {author} {\bibfnamefont {G.}~\bibnamefont
  {Panotopoulos}}\ and\ \bibinfo {author} {\bibfnamefont {I.}~\bibnamefont
  {Lopes}},\ }\href {\doibase 10.1142/S0218271820500583} {\bibfield  {journal}
  {\bibinfo  {journal} {Int. J. Mod. Phys. D}\ }\textbf {\bibinfo {volume}
  {29}},\ \bibinfo {pages} {2050058} (\bibinfo {year} {2020})},\ \Eprint
  {http://arxiv.org/abs/2005.11563} {arXiv:2005.11563 [hep-ph]} \BibitemShut
  {NoStop}%
\bibitem [{\citenamefont {Dasgupta}\ \emph {et~al.}(2020)\citenamefont
  {Dasgupta}, \citenamefont {Gupta},\ and\ \citenamefont
  {Ray}}]{Dasgupta:2020dik}%
  \BibitemOpen
  \bibfield  {author} {\bibinfo {author} {\bibfnamefont {B.}~\bibnamefont
  {Dasgupta}}, \bibinfo {author} {\bibfnamefont {A.}~\bibnamefont {Gupta}}, \
  and\ \bibinfo {author} {\bibfnamefont {A.}~\bibnamefont {Ray}},\ }\href
  {\doibase 10.1088/1475-7516/2020/10/023} {\bibfield  {journal} {\bibinfo
  {journal} {JCAP}\ }\textbf {\bibinfo {volume} {10}},\ \bibinfo {pages} {023}
  (\bibinfo {year} {2020})},\ \Eprint {http://arxiv.org/abs/2006.10773}
  {arXiv:2006.10773 [hep-ph]} \BibitemShut {NoStop}%
\bibitem [{\citenamefont {Horowitz}(2020)}]{Horowitz:2020axx}%
  \BibitemOpen
  \bibfield  {author} {\bibinfo {author} {\bibfnamefont {C.~J.}\ \bibnamefont
  {Horowitz}},\ }\href {\doibase 10.1103/PhysRevD.102.083031} {\bibfield
  {journal} {\bibinfo  {journal} {Phys. Rev. D}\ }\textbf {\bibinfo {volume}
  {102}},\ \bibinfo {pages} {083031} (\bibinfo {year} {2020})},\ \Eprint
  {http://arxiv.org/abs/2008.03291} {arXiv:2008.03291 [astro-ph.SR]}
  \BibitemShut {NoStop}%
\bibitem [{\citenamefont {Garani}\ \emph {et~al.}(2021)\citenamefont {Garani},
  \citenamefont {Gupta},\ and\ \citenamefont {Raj}}]{Garani:2020wge}%
  \BibitemOpen
  \bibfield  {author} {\bibinfo {author} {\bibfnamefont {R.}~\bibnamefont
  {Garani}}, \bibinfo {author} {\bibfnamefont {A.}~\bibnamefont {Gupta}}, \
  and\ \bibinfo {author} {\bibfnamefont {N.}~\bibnamefont {Raj}},\ }\href
  {\doibase 10.1103/PhysRevD.103.043019} {\bibfield  {journal} {\bibinfo
  {journal} {Phys. Rev. D}\ }\textbf {\bibinfo {volume} {103}},\ \bibinfo
  {pages} {043019} (\bibinfo {year} {2021})},\ \Eprint
  {http://arxiv.org/abs/2009.10728} {arXiv:2009.10728 [hep-ph]} \BibitemShut
  {NoStop}%
\bibitem [{\citenamefont {Bose}\ and\ \citenamefont
  {Sarkar}(2023)}]{Bose:2022ola}%
  \BibitemOpen
  \bibfield  {author} {\bibinfo {author} {\bibfnamefont {D.}~\bibnamefont
  {Bose}}\ and\ \bibinfo {author} {\bibfnamefont {S.}~\bibnamefont {Sarkar}},\
  }\href {\doibase 10.1103/PhysRevD.107.063010} {\bibfield  {journal} {\bibinfo
   {journal} {Phys. Rev. D}\ }\textbf {\bibinfo {volume} {107}},\ \bibinfo
  {pages} {063010} (\bibinfo {year} {2023})},\ \Eprint
  {http://arxiv.org/abs/2211.16982} {arXiv:2211.16982 [astro-ph.CO]}
  \BibitemShut {NoStop}%
\bibitem [{\citenamefont {Hardy}\ and\ \citenamefont
  {Song}(2023)}]{Hardy:2022ufh}%
  \BibitemOpen
  \bibfield  {author} {\bibinfo {author} {\bibfnamefont {E.}~\bibnamefont
  {Hardy}}\ and\ \bibinfo {author} {\bibfnamefont {N.}~\bibnamefont {Song}},\
  }\href {\doibase 10.1103/PhysRevD.107.115035} {\bibfield  {journal} {\bibinfo
   {journal} {Phys. Rev. D}\ }\textbf {\bibinfo {volume} {107}},\ \bibinfo
  {pages} {115035} (\bibinfo {year} {2023})},\ \Eprint
  {http://arxiv.org/abs/2212.09756} {arXiv:2212.09756 [hep-ph]} \BibitemShut
  {NoStop}%
\bibitem [{\citenamefont {Nguyen}\ and\ \citenamefont
  {Tait}(2023)}]{Nguyen:2022zwb}%
  \BibitemOpen
  \bibfield  {author} {\bibinfo {author} {\bibfnamefont {T.~T.~Q.}\
  \bibnamefont {Nguyen}}\ and\ \bibinfo {author} {\bibfnamefont {T.~M.~P.}\
  \bibnamefont {Tait}},\ }\href {\doibase 10.1103/PhysRevD.107.115016}
  {\bibfield  {journal} {\bibinfo  {journal} {Phys. Rev. D}\ }\textbf {\bibinfo
  {volume} {107}},\ \bibinfo {pages} {115016} (\bibinfo {year} {2023})},\
  \Eprint {http://arxiv.org/abs/2212.12547} {arXiv:2212.12547 [hep-ph]}
  \BibitemShut {NoStop}%
\bibitem [{\citenamefont {Linden}\ \emph {et~al.}(2024)\citenamefont {Linden},
  \citenamefont {Nguyen},\ and\ \citenamefont {Tait}}]{Linden:2024uph}%
  \BibitemOpen
  \bibfield  {author} {\bibinfo {author} {\bibfnamefont {T.}~\bibnamefont
  {Linden}}, \bibinfo {author} {\bibfnamefont {T.~T.~Q.}\ \bibnamefont
  {Nguyen}}, \ and\ \bibinfo {author} {\bibfnamefont {T.~M.~P.}\ \bibnamefont
  {Tait}},\ }\href@noop {} {\  (\bibinfo {year} {2024})},\ \Eprint
  {http://arxiv.org/abs/2402.01839} {arXiv:2402.01839 [hep-ph]} \BibitemShut
  {NoStop}%
\bibitem [{\citenamefont {Song}\ \emph {et~al.}(2024)\citenamefont {Song},
  \citenamefont {Su},\ and\ \citenamefont {Wu}}]{Song:2024rru}%
  \BibitemOpen
  \bibfield  {author} {\bibinfo {author} {\bibfnamefont {N.}~\bibnamefont
  {Song}}, \bibinfo {author} {\bibfnamefont {L.}~\bibnamefont {Su}}, \ and\
  \bibinfo {author} {\bibfnamefont {L.}~\bibnamefont {Wu}},\ }\href@noop {} {\
  (\bibinfo {year} {2024})},\ \Eprint {http://arxiv.org/abs/2402.15144}
  {arXiv:2402.15144 [hep-ph]} \BibitemShut {NoStop}%
\bibitem [{\citenamefont {Yadav}\ \emph {et~al.}(2024)\citenamefont {Yadav},
  \citenamefont {Mishra},\ and\ \citenamefont {Sarkar}}]{Yadav:2024xob}%
  \BibitemOpen
  \bibfield  {author} {\bibinfo {author} {\bibfnamefont {S.}~\bibnamefont
  {Yadav}}, \bibinfo {author} {\bibfnamefont {M.}~\bibnamefont {Mishra}}, \
  and\ \bibinfo {author} {\bibfnamefont {T.~G.}\ \bibnamefont {Sarkar}},\
  }\href@noop {} {\  (\bibinfo {year} {2024})},\ \Eprint
  {http://arxiv.org/abs/2403.15305} {arXiv:2403.15305 [astro-ph.HE]}
  \BibitemShut {NoStop}%
\bibitem [{\citenamefont {Lu}\ \emph {et~al.}(2024)\citenamefont {Lu},
  \citenamefont {Mishra},\ and\ \citenamefont {Wu}}]{Lu:2024kiz}%
  \BibitemOpen
  \bibfield  {author} {\bibinfo {author} {\bibfnamefont {C.-T.}\ \bibnamefont
  {Lu}}, \bibinfo {author} {\bibfnamefont {A.~K.}\ \bibnamefont {Mishra}}, \
  and\ \bibinfo {author} {\bibfnamefont {L.}~\bibnamefont {Wu}},\ }\href@noop
  {} {\  (\bibinfo {year} {2024})},\ \Eprint {http://arxiv.org/abs/2404.07187}
  {arXiv:2404.07187 [hep-ph]} \BibitemShut {NoStop}%
\bibitem [{\citenamefont {Ema}\ \emph {et~al.}(2024)\citenamefont {Ema},
  \citenamefont {McGehee}, \citenamefont {Pospelov},\ and\ \citenamefont
  {Ray}}]{Ema:2024wqr}%
  \BibitemOpen
  \bibfield  {author} {\bibinfo {author} {\bibfnamefont {Y.}~\bibnamefont
  {Ema}}, \bibinfo {author} {\bibfnamefont {R.}~\bibnamefont {McGehee}},
  \bibinfo {author} {\bibfnamefont {M.}~\bibnamefont {Pospelov}}, \ and\
  \bibinfo {author} {\bibfnamefont {A.}~\bibnamefont {Ray}},\ }\href@noop {} {\
   (\bibinfo {year} {2024})},\ \Eprint {http://arxiv.org/abs/2405.18472}
  {arXiv:2405.18472 [hep-ph]} \BibitemShut {NoStop}%
\bibitem [{\citenamefont {Liu}\ and\ \citenamefont
  {Mishra}(2024)}]{Liu:2024qbe}%
  \BibitemOpen
  \bibfield  {author} {\bibinfo {author} {\bibfnamefont {N.}~\bibnamefont
  {Liu}}\ and\ \bibinfo {author} {\bibfnamefont {A.~K.}\ \bibnamefont
  {Mishra}},\ }\href@noop {} {\  (\bibinfo {year} {2024})},\ \Eprint
  {http://arxiv.org/abs/2408.00594} {arXiv:2408.00594 [astro-ph.CO]}
  \BibitemShut {NoStop}%
\bibitem [{\citenamefont {Das}\ \emph {et~al.}(2024)\citenamefont {Das},
  \citenamefont {Dev}, \citenamefont {Okawa},\ and\ \citenamefont
  {Soni}}]{Das:2024thc}%
  \BibitemOpen
  \bibfield  {author} {\bibinfo {author} {\bibfnamefont {S.}~\bibnamefont
  {Das}}, \bibinfo {author} {\bibfnamefont {P.~S.~B.}\ \bibnamefont {Dev}},
  \bibinfo {author} {\bibfnamefont {T.}~\bibnamefont {Okawa}}, \ and\ \bibinfo
  {author} {\bibfnamefont {A.}~\bibnamefont {Soni}},\ }\href@noop {} {\
  (\bibinfo {year} {2024})},\ \Eprint {http://arxiv.org/abs/2408.01484}
  {arXiv:2408.01484 [hep-ph]} \BibitemShut {NoStop}%
\bibitem [{\citenamefont {Goldman}\ and\ \citenamefont
  {Nussinov}(1989)}]{Goldman:1989nd}%
  \BibitemOpen
  \bibfield  {author} {\bibinfo {author} {\bibfnamefont {I.}~\bibnamefont
  {Goldman}}\ and\ \bibinfo {author} {\bibfnamefont {S.}~\bibnamefont
  {Nussinov}},\ }\href {\doibase 10.1103/PhysRevD.40.3221} {\bibfield
  {journal} {\bibinfo  {journal} {Phys. Rev. D}\ }\textbf {\bibinfo {volume}
  {40}},\ \bibinfo {pages} {3221} (\bibinfo {year} {1989})}\BibitemShut
  {NoStop}%
\bibitem [{\citenamefont {Kouvaris}\ and\ \citenamefont
  {Tinyakov}(2010)}]{Kouvaris:2010vv}%
  \BibitemOpen
  \bibfield  {author} {\bibinfo {author} {\bibfnamefont {C.}~\bibnamefont
  {Kouvaris}}\ and\ \bibinfo {author} {\bibfnamefont {P.}~\bibnamefont
  {Tinyakov}},\ }\href {\doibase 10.1103/PhysRevD.82.063531} {\bibfield
  {journal} {\bibinfo  {journal} {Phys. Rev. D}\ }\textbf {\bibinfo {volume}
  {82}},\ \bibinfo {pages} {063531} (\bibinfo {year} {2010})},\ \Eprint
  {http://arxiv.org/abs/1004.0586} {arXiv:1004.0586 [astro-ph.GA]} \BibitemShut
  {NoStop}%
\bibitem [{\citenamefont {McDermott}\ \emph {et~al.}(2012)\citenamefont
  {McDermott}, \citenamefont {Yu},\ and\ \citenamefont
  {Zurek}}]{McDermott:2011jp}%
  \BibitemOpen
  \bibfield  {author} {\bibinfo {author} {\bibfnamefont {S.~D.}\ \bibnamefont
  {McDermott}}, \bibinfo {author} {\bibfnamefont {H.-B.}\ \bibnamefont {Yu}}, \
  and\ \bibinfo {author} {\bibfnamefont {K.~M.}\ \bibnamefont {Zurek}},\ }\href
  {\doibase 10.1103/PhysRevD.85.023519} {\bibfield  {journal} {\bibinfo
  {journal} {Phys. Rev. D}\ }\textbf {\bibinfo {volume} {85}},\ \bibinfo
  {pages} {023519} (\bibinfo {year} {2012})},\ \Eprint
  {http://arxiv.org/abs/1103.5472} {arXiv:1103.5472 [hep-ph]} \BibitemShut
  {NoStop}%
\bibitem [{\citenamefont {Kouvaris}\ and\ \citenamefont
  {Tinyakov}(2011)}]{Kouvaris:2011fi}%
  \BibitemOpen
  \bibfield  {author} {\bibinfo {author} {\bibfnamefont {C.}~\bibnamefont
  {Kouvaris}}\ and\ \bibinfo {author} {\bibfnamefont {P.}~\bibnamefont
  {Tinyakov}},\ }\href {\doibase 10.1103/PhysRevLett.107.091301} {\bibfield
  {journal} {\bibinfo  {journal} {Phys. Rev. Lett.}\ }\textbf {\bibinfo
  {volume} {107}},\ \bibinfo {pages} {091301} (\bibinfo {year} {2011})},\
  \Eprint {http://arxiv.org/abs/1104.0382} {arXiv:1104.0382 [astro-ph.CO]}
  \BibitemShut {NoStop}%
\bibitem [{\citenamefont {Bramante}\ and\ \citenamefont
  {Linden}(2014)}]{Bramante:2014zca}%
  \BibitemOpen
  \bibfield  {author} {\bibinfo {author} {\bibfnamefont {J.}~\bibnamefont
  {Bramante}}\ and\ \bibinfo {author} {\bibfnamefont {T.}~\bibnamefont
  {Linden}},\ }\href {\doibase 10.1103/PhysRevLett.113.191301} {\bibfield
  {journal} {\bibinfo  {journal} {Phys. Rev. Lett.}\ }\textbf {\bibinfo
  {volume} {113}},\ \bibinfo {pages} {191301} (\bibinfo {year} {2014})},\
  \Eprint {http://arxiv.org/abs/1405.1031} {arXiv:1405.1031 [astro-ph.HE]}
  \BibitemShut {NoStop}%
\bibitem [{\citenamefont {Fuller}\ and\ \citenamefont
  {Ott}(2015)}]{Fuller:2014rza}%
  \BibitemOpen
  \bibfield  {author} {\bibinfo {author} {\bibfnamefont {J.}~\bibnamefont
  {Fuller}}\ and\ \bibinfo {author} {\bibfnamefont {C.}~\bibnamefont {Ott}},\
  }\href {\doibase 10.1093/mnrasl/slv049} {\bibfield  {journal} {\bibinfo
  {journal} {Mon. Not. Roy. Astron. Soc.}\ }\textbf {\bibinfo {volume} {450}},\
  \bibinfo {pages} {L71} (\bibinfo {year} {2015})},\ \Eprint
  {http://arxiv.org/abs/1412.6119} {arXiv:1412.6119 [astro-ph.HE]} \BibitemShut
  {NoStop}%
\bibitem [{\citenamefont {Bramante}\ \emph {et~al.}(2018)\citenamefont
  {Bramante}, \citenamefont {Linden},\ and\ \citenamefont
  {Tsai}}]{Bramante:2017ulk}%
  \BibitemOpen
  \bibfield  {author} {\bibinfo {author} {\bibfnamefont {J.}~\bibnamefont
  {Bramante}}, \bibinfo {author} {\bibfnamefont {T.}~\bibnamefont {Linden}}, \
  and\ \bibinfo {author} {\bibfnamefont {Y.-D.}\ \bibnamefont {Tsai}},\ }\href
  {\doibase 10.1103/PhysRevD.97.055016} {\bibfield  {journal} {\bibinfo
  {journal} {Phys. Rev. D}\ }\textbf {\bibinfo {volume} {97}},\ \bibinfo
  {pages} {055016} (\bibinfo {year} {2018})},\ \Eprint
  {http://arxiv.org/abs/1706.00001} {arXiv:1706.00001 [hep-ph]} \BibitemShut
  {NoStop}%
\bibitem [{\citenamefont {Garani}\ \emph {et~al.}(2019)\citenamefont {Garani},
  \citenamefont {Genolini},\ and\ \citenamefont {Hambye}}]{Garani:2018kkd}%
  \BibitemOpen
  \bibfield  {author} {\bibinfo {author} {\bibfnamefont {R.}~\bibnamefont
  {Garani}}, \bibinfo {author} {\bibfnamefont {Y.}~\bibnamefont {Genolini}}, \
  and\ \bibinfo {author} {\bibfnamefont {T.}~\bibnamefont {Hambye}},\ }\href
  {\doibase 10.1088/1475-7516/2019/05/035} {\bibfield  {journal} {\bibinfo
  {journal} {JCAP}\ }\textbf {\bibinfo {volume} {05}},\ \bibinfo {pages} {035}
  (\bibinfo {year} {2019})},\ \Eprint {http://arxiv.org/abs/1812.08773}
  {arXiv:1812.08773 [hep-ph]} \BibitemShut {NoStop}%
\bibitem [{\citenamefont {Dasgupta}\ \emph {et~al.}(2021)\citenamefont
  {Dasgupta}, \citenamefont {Laha},\ and\ \citenamefont
  {Ray}}]{Dasgupta:2020mqg}%
  \BibitemOpen
  \bibfield  {author} {\bibinfo {author} {\bibfnamefont {B.}~\bibnamefont
  {Dasgupta}}, \bibinfo {author} {\bibfnamefont {R.}~\bibnamefont {Laha}}, \
  and\ \bibinfo {author} {\bibfnamefont {A.}~\bibnamefont {Ray}},\ }\href
  {\doibase 10.1103/PhysRevLett.126.141105} {\bibfield  {journal} {\bibinfo
  {journal} {Phys. Rev. Lett.}\ }\textbf {\bibinfo {volume} {126}},\ \bibinfo
  {pages} {141105} (\bibinfo {year} {2021})},\ \Eprint
  {http://arxiv.org/abs/2009.01825} {arXiv:2009.01825 [astro-ph.HE]}
  \BibitemShut {NoStop}%
\bibitem [{\citenamefont {Giffin}\ \emph {et~al.}(2022)\citenamefont {Giffin},
  \citenamefont {Lloyd}, \citenamefont {McDermott},\ and\ \citenamefont
  {Profumo}}]{Giffin:2021kgb}%
  \BibitemOpen
  \bibfield  {author} {\bibinfo {author} {\bibfnamefont {P.}~\bibnamefont
  {Giffin}}, \bibinfo {author} {\bibfnamefont {J.}~\bibnamefont {Lloyd}},
  \bibinfo {author} {\bibfnamefont {S.~D.}\ \bibnamefont {McDermott}}, \ and\
  \bibinfo {author} {\bibfnamefont {S.}~\bibnamefont {Profumo}},\ }\href
  {\doibase 10.1103/PhysRevD.105.123030} {\bibfield  {journal} {\bibinfo
  {journal} {Phys. Rev. D}\ }\textbf {\bibinfo {volume} {105}},\ \bibinfo
  {pages} {123030} (\bibinfo {year} {2022})},\ \Eprint
  {http://arxiv.org/abs/2105.06504} {arXiv:2105.06504 [hep-ph]} \BibitemShut
  {NoStop}%
\bibitem [{\citenamefont {Garani}\ \emph {et~al.}(2022)\citenamefont {Garani},
  \citenamefont {Levkov},\ and\ \citenamefont {Tinyakov}}]{Garani:2021gvc}%
  \BibitemOpen
  \bibfield  {author} {\bibinfo {author} {\bibfnamefont {R.}~\bibnamefont
  {Garani}}, \bibinfo {author} {\bibfnamefont {D.}~\bibnamefont {Levkov}}, \
  and\ \bibinfo {author} {\bibfnamefont {P.}~\bibnamefont {Tinyakov}},\ }\href
  {\doibase 10.1103/PhysRevD.105.063019} {\bibfield  {journal} {\bibinfo
  {journal} {Phys. Rev. D}\ }\textbf {\bibinfo {volume} {105}},\ \bibinfo
  {pages} {063019} (\bibinfo {year} {2022})},\ \Eprint
  {http://arxiv.org/abs/2112.09716} {arXiv:2112.09716 [hep-ph]} \BibitemShut
  {NoStop}%
\bibitem [{\citenamefont {Bhattacharya}\ \emph {et~al.}(2023)\citenamefont
  {Bhattacharya}, \citenamefont {Dasgupta}, \citenamefont {Laha},\ and\
  \citenamefont {Ray}}]{Bhattacharya:2023stq}%
  \BibitemOpen
  \bibfield  {author} {\bibinfo {author} {\bibfnamefont {S.}~\bibnamefont
  {Bhattacharya}}, \bibinfo {author} {\bibfnamefont {B.}~\bibnamefont
  {Dasgupta}}, \bibinfo {author} {\bibfnamefont {R.}~\bibnamefont {Laha}}, \
  and\ \bibinfo {author} {\bibfnamefont {A.}~\bibnamefont {Ray}},\ }\href
  {\doibase 10.1103/PhysRevLett.131.091401} {\bibfield  {journal} {\bibinfo
  {journal} {Phys. Rev. Lett.}\ }\textbf {\bibinfo {volume} {131}},\ \bibinfo
  {pages} {091401} (\bibinfo {year} {2023})},\ \Eprint
  {http://arxiv.org/abs/2302.07898} {arXiv:2302.07898 [hep-ph]} \BibitemShut
  {NoStop}%
\bibitem [{\citenamefont {Kouvaris}(2008)}]{Kouvaris:2007ay}%
  \BibitemOpen
  \bibfield  {author} {\bibinfo {author} {\bibfnamefont {C.}~\bibnamefont
  {Kouvaris}},\ }\href {\doibase 10.1103/PhysRevD.77.023006} {\bibfield
  {journal} {\bibinfo  {journal} {Phys. Rev. D}\ }\textbf {\bibinfo {volume}
  {77}},\ \bibinfo {pages} {023006} (\bibinfo {year} {2008})},\ \Eprint
  {http://arxiv.org/abs/0708.2362} {arXiv:0708.2362 [astro-ph]} \BibitemShut
  {NoStop}%
\bibitem [{\citenamefont {Baryakhtar}\ \emph {et~al.}(2017)\citenamefont
  {Baryakhtar}, \citenamefont {Bramante}, \citenamefont {Li}, \citenamefont
  {Linden},\ and\ \citenamefont {Raj}}]{Baryakhtar:2017dbj}%
  \BibitemOpen
  \bibfield  {author} {\bibinfo {author} {\bibfnamefont {M.}~\bibnamefont
  {Baryakhtar}}, \bibinfo {author} {\bibfnamefont {J.}~\bibnamefont
  {Bramante}}, \bibinfo {author} {\bibfnamefont {S.~W.}\ \bibnamefont {Li}},
  \bibinfo {author} {\bibfnamefont {T.}~\bibnamefont {Linden}}, \ and\ \bibinfo
  {author} {\bibfnamefont {N.}~\bibnamefont {Raj}},\ }\href {\doibase
  10.1103/PhysRevLett.119.131801} {\bibfield  {journal} {\bibinfo  {journal}
  {Phys. Rev. Lett.}\ }\textbf {\bibinfo {volume} {119}},\ \bibinfo {pages}
  {131801} (\bibinfo {year} {2017})},\ \Eprint
  {http://arxiv.org/abs/1704.01577} {arXiv:1704.01577 [hep-ph]} \BibitemShut
  {NoStop}%
\bibitem [{\citenamefont {Raj}\ \emph {et~al.}(2018)\citenamefont {Raj},
  \citenamefont {Tanedo},\ and\ \citenamefont {Yu}}]{Raj:2017wrv}%
  \BibitemOpen
  \bibfield  {author} {\bibinfo {author} {\bibfnamefont {N.}~\bibnamefont
  {Raj}}, \bibinfo {author} {\bibfnamefont {P.}~\bibnamefont {Tanedo}}, \ and\
  \bibinfo {author} {\bibfnamefont {H.-B.}\ \bibnamefont {Yu}},\ }\href
  {\doibase 10.1103/PhysRevD.97.043006} {\bibfield  {journal} {\bibinfo
  {journal} {Phys. Rev. D}\ }\textbf {\bibinfo {volume} {97}},\ \bibinfo
  {pages} {043006} (\bibinfo {year} {2018})},\ \Eprint
  {http://arxiv.org/abs/1707.09442} {arXiv:1707.09442 [hep-ph]} \BibitemShut
  {NoStop}%
\bibitem [{\citenamefont {Camargo}\ \emph {et~al.}(2019)\citenamefont
  {Camargo}, \citenamefont {Queiroz},\ and\ \citenamefont
  {Sturani}}]{Camargo:2019wou}%
  \BibitemOpen
  \bibfield  {author} {\bibinfo {author} {\bibfnamefont {D.~A.}\ \bibnamefont
  {Camargo}}, \bibinfo {author} {\bibfnamefont {F.~S.}\ \bibnamefont
  {Queiroz}}, \ and\ \bibinfo {author} {\bibfnamefont {R.}~\bibnamefont
  {Sturani}},\ }\href {\doibase 10.1088/1475-7516/2019/09/051} {\bibfield
  {journal} {\bibinfo  {journal} {JCAP}\ }\textbf {\bibinfo {volume} {09}},\
  \bibinfo {pages} {051} (\bibinfo {year} {2019})},\ \Eprint
  {http://arxiv.org/abs/1901.05474} {arXiv:1901.05474 [hep-ph]} \BibitemShut
  {NoStop}%
\bibitem [{\citenamefont {Bell}\ \emph {et~al.}(2019)\citenamefont {Bell},
  \citenamefont {Busoni},\ and\ \citenamefont {Robles}}]{Bell:2019pyc}%
  \BibitemOpen
  \bibfield  {author} {\bibinfo {author} {\bibfnamefont {N.~F.}\ \bibnamefont
  {Bell}}, \bibinfo {author} {\bibfnamefont {G.}~\bibnamefont {Busoni}}, \ and\
  \bibinfo {author} {\bibfnamefont {S.}~\bibnamefont {Robles}},\ }\href
  {\doibase 10.1088/1475-7516/2019/06/054} {\bibfield  {journal} {\bibinfo
  {journal} {JCAP}\ }\textbf {\bibinfo {volume} {06}},\ \bibinfo {pages} {054}
  (\bibinfo {year} {2019})},\ \Eprint {http://arxiv.org/abs/1904.09803}
  {arXiv:1904.09803 [hep-ph]} \BibitemShut {NoStop}%
\bibitem [{\citenamefont {Hamaguchi}\ \emph {et~al.}(2019)\citenamefont
  {Hamaguchi}, \citenamefont {Nagata},\ and\ \citenamefont
  {Yanagi}}]{Hamaguchi:2019oev}%
  \BibitemOpen
  \bibfield  {author} {\bibinfo {author} {\bibfnamefont {K.}~\bibnamefont
  {Hamaguchi}}, \bibinfo {author} {\bibfnamefont {N.}~\bibnamefont {Nagata}}, \
  and\ \bibinfo {author} {\bibfnamefont {K.}~\bibnamefont {Yanagi}},\ }\href
  {\doibase 10.1016/j.physletb.2019.06.060} {\bibfield  {journal} {\bibinfo
  {journal} {Phys. Lett. B}\ }\textbf {\bibinfo {volume} {795}},\ \bibinfo
  {pages} {484} (\bibinfo {year} {2019})},\ \Eprint
  {http://arxiv.org/abs/1905.02991} {arXiv:1905.02991 [hep-ph]} \BibitemShut
  {NoStop}%
\bibitem [{\citenamefont {Garani}\ and\ \citenamefont
  {Heeck}(2019)}]{Garani:2019fpa}%
  \BibitemOpen
  \bibfield  {author} {\bibinfo {author} {\bibfnamefont {R.}~\bibnamefont
  {Garani}}\ and\ \bibinfo {author} {\bibfnamefont {J.}~\bibnamefont {Heeck}},\
  }\href {\doibase 10.1103/PhysRevD.100.035039} {\bibfield  {journal} {\bibinfo
   {journal} {Phys. Rev. D}\ }\textbf {\bibinfo {volume} {100}},\ \bibinfo
  {pages} {035039} (\bibinfo {year} {2019})},\ \Eprint
  {http://arxiv.org/abs/1906.10145} {arXiv:1906.10145 [hep-ph]} \BibitemShut
  {NoStop}%
\bibitem [{\citenamefont {Acevedo}\ \emph {et~al.}(2020)\citenamefont
  {Acevedo}, \citenamefont {Bramante}, \citenamefont {Leane},\ and\
  \citenamefont {Raj}}]{Acevedo:2019agu}%
  \BibitemOpen
  \bibfield  {author} {\bibinfo {author} {\bibfnamefont {J.~F.}\ \bibnamefont
  {Acevedo}}, \bibinfo {author} {\bibfnamefont {J.}~\bibnamefont {Bramante}},
  \bibinfo {author} {\bibfnamefont {R.~K.}\ \bibnamefont {Leane}}, \ and\
  \bibinfo {author} {\bibfnamefont {N.}~\bibnamefont {Raj}},\ }\href {\doibase
  10.1088/1475-7516/2020/03/038} {\bibfield  {journal} {\bibinfo  {journal}
  {JCAP}\ }\textbf {\bibinfo {volume} {03}},\ \bibinfo {pages} {038} (\bibinfo
  {year} {2020})},\ \Eprint {http://arxiv.org/abs/1911.06334} {arXiv:1911.06334
  [hep-ph]} \BibitemShut {NoStop}%
\bibitem [{\citenamefont {Joglekar}\ \emph
  {et~al.}(2020{\natexlab{a}})\citenamefont {Joglekar}, \citenamefont {Raj},
  \citenamefont {Tanedo},\ and\ \citenamefont {Yu}}]{Joglekar:2019vzy}%
  \BibitemOpen
  \bibfield  {author} {\bibinfo {author} {\bibfnamefont {A.}~\bibnamefont
  {Joglekar}}, \bibinfo {author} {\bibfnamefont {N.}~\bibnamefont {Raj}},
  \bibinfo {author} {\bibfnamefont {P.}~\bibnamefont {Tanedo}}, \ and\ \bibinfo
  {author} {\bibfnamefont {H.-B.}\ \bibnamefont {Yu}},\ }\href {\doibase
  10.1016/j.physletb.2020.135767} {\bibfield  {journal} {\bibinfo  {journal}
  {Phys. Lett. B}\ }\textbf {\bibinfo {volume} {809}},\ \bibinfo {pages}
  {135767} (\bibinfo {year} {2020}{\natexlab{a}})},\ \Eprint
  {http://arxiv.org/abs/1911.13293} {arXiv:1911.13293 [hep-ph]} \BibitemShut
  {NoStop}%
\bibitem [{\citenamefont {Joglekar}\ \emph
  {et~al.}(2020{\natexlab{b}})\citenamefont {Joglekar}, \citenamefont {Raj},
  \citenamefont {Tanedo},\ and\ \citenamefont {Yu}}]{Joglekar:2020liw}%
  \BibitemOpen
  \bibfield  {author} {\bibinfo {author} {\bibfnamefont {A.}~\bibnamefont
  {Joglekar}}, \bibinfo {author} {\bibfnamefont {N.}~\bibnamefont {Raj}},
  \bibinfo {author} {\bibfnamefont {P.}~\bibnamefont {Tanedo}}, \ and\ \bibinfo
  {author} {\bibfnamefont {H.-B.}\ \bibnamefont {Yu}},\ }\href {\doibase
  10.1103/PhysRevD.102.123002} {\bibfield  {journal} {\bibinfo  {journal}
  {Phys. Rev. D}\ }\textbf {\bibinfo {volume} {102}},\ \bibinfo {pages}
  {123002} (\bibinfo {year} {2020}{\natexlab{b}})},\ \Eprint
  {http://arxiv.org/abs/2004.09539} {arXiv:2004.09539 [hep-ph]} \BibitemShut
  {NoStop}%
\bibitem [{\citenamefont {Page}\ \emph {et~al.}(2004)\citenamefont {Page},
  \citenamefont {Lattimer}, \citenamefont {Prakash},\ and\ \citenamefont
  {Steiner}}]{Page:2004fy}%
  \BibitemOpen
  \bibfield  {author} {\bibinfo {author} {\bibfnamefont {D.}~\bibnamefont
  {Page}}, \bibinfo {author} {\bibfnamefont {J.~M.}\ \bibnamefont {Lattimer}},
  \bibinfo {author} {\bibfnamefont {M.}~\bibnamefont {Prakash}}, \ and\
  \bibinfo {author} {\bibfnamefont {A.~W.}\ \bibnamefont {Steiner}},\ }\href
  {\doibase 10.1086/424844} {\bibfield  {journal} {\bibinfo  {journal}
  {Astrophys. J. Suppl.}\ }\textbf {\bibinfo {volume} {155}},\ \bibinfo {pages}
  {623} (\bibinfo {year} {2004})},\ \Eprint
  {http://arxiv.org/abs/astro-ph/0403657} {arXiv:astro-ph/0403657} \BibitemShut
  {NoStop}%
\bibitem [{\citenamefont {Yakovlev}\ \emph {et~al.}(2005)\citenamefont
  {Yakovlev}, \citenamefont {Gnedin}, \citenamefont {Gusakov}, \citenamefont
  {Kaminker}, \citenamefont {Levenfish},\ and\ \citenamefont
  {Potekhin}}]{Yakovlev:2004yr}%
  \BibitemOpen
  \bibfield  {author} {\bibinfo {author} {\bibfnamefont {D.~G.}\ \bibnamefont
  {Yakovlev}}, \bibinfo {author} {\bibfnamefont {O.~Y.}\ \bibnamefont
  {Gnedin}}, \bibinfo {author} {\bibfnamefont {M.~E.}\ \bibnamefont {Gusakov}},
  \bibinfo {author} {\bibfnamefont {A.~D.}\ \bibnamefont {Kaminker}}, \bibinfo
  {author} {\bibfnamefont {K.~P.}\ \bibnamefont {Levenfish}}, \ and\ \bibinfo
  {author} {\bibfnamefont {A.~Y.}\ \bibnamefont {Potekhin}},\ }\href {\doibase
  10.1016/j.nuclphysa.2005.02.061} {\bibfield  {journal} {\bibinfo  {journal}
  {Nucl. Phys. A}\ }\textbf {\bibinfo {volume} {752}},\ \bibinfo {pages} {590}
  (\bibinfo {year} {2005})},\ \Eprint {http://arxiv.org/abs/astro-ph/0409751}
  {arXiv:astro-ph/0409751} \BibitemShut {NoStop}%
\bibitem [{\citenamefont {Gardner}\ \emph {et~al.}(2006)\citenamefont {Gardner}
  \emph {et~al.}}]{Gardner:2006ky}%
  \BibitemOpen
  \bibfield  {author} {\bibinfo {author} {\bibfnamefont {J.~P.}\ \bibnamefont
  {Gardner}} \emph {et~al.},\ }\href {\doibase 10.1007/s11214-006-8315-7}
  {\bibfield  {journal} {\bibinfo  {journal} {Space Sci. Rev.}\ }\textbf
  {\bibinfo {volume} {123}},\ \bibinfo {pages} {485} (\bibinfo {year}
  {2006})},\ \Eprint {http://arxiv.org/abs/astro-ph/0606175}
  {arXiv:astro-ph/0606175} \BibitemShut {NoStop}%
\bibitem [{\citenamefont {Kalirai}(2018)}]{Kalirai:2018qfg}%
  \BibitemOpen
  \bibfield  {author} {\bibinfo {author} {\bibfnamefont {J.}~\bibnamefont
  {Kalirai}},\ }\href {\doibase 10.1080/00107514.2018.1467648} {\bibfield
  {journal} {\bibinfo  {journal} {Contemp. Phys.}\ }\textbf {\bibinfo {volume}
  {59}},\ \bibinfo {pages} {251} (\bibinfo {year} {2018})},\ \Eprint
  {http://arxiv.org/abs/1805.06941} {arXiv:1805.06941 [astro-ph.IM]}
  \BibitemShut {NoStop}%
\bibitem [{\citenamefont {Skidmore}\ \emph {et~al.}(2015)\citenamefont
  {Skidmore} \emph
  {et~al.}}]{TMTInternationalScienceDevelopmentTeamsTMTScienceAdvisoryCommittee:2015pvw}%
  \BibitemOpen
  \bibfield  {author} {\bibinfo {author} {\bibfnamefont {W.}~\bibnamefont
  {Skidmore}} \emph {et~al.} (\bibinfo {collaboration} {TMT International
  Science Development Teams \& TMT Science Advisory Committee}),\ }\href
  {\doibase 10.1088/1674-4527/15/12/001} {\bibfield  {journal} {\bibinfo
  {journal} {Res. Astron. Astrophys.}\ }\textbf {\bibinfo {volume} {15}},\
  \bibinfo {pages} {1945} (\bibinfo {year} {2015})},\ \Eprint
  {http://arxiv.org/abs/1505.01195} {arXiv:1505.01195 [astro-ph.IM]}
  \BibitemShut {NoStop}%
\bibitem [{\citenamefont {Neichel}\ \emph {et~al.}(2018)\citenamefont
  {Neichel}, \citenamefont {Mouillet}, \citenamefont {Gendron}, \citenamefont
  {Correia}, \citenamefont {Sauvage},\ and\ \citenamefont
  {Fusco}}]{neichel2018overvieweuropeanextremelylarge}%
  \BibitemOpen
  \bibfield  {author} {\bibinfo {author} {\bibfnamefont {B.}~\bibnamefont
  {Neichel}}, \bibinfo {author} {\bibfnamefont {D.}~\bibnamefont {Mouillet}},
  \bibinfo {author} {\bibfnamefont {E.}~\bibnamefont {Gendron}}, \bibinfo
  {author} {\bibfnamefont {C.}~\bibnamefont {Correia}}, \bibinfo {author}
  {\bibfnamefont {J.~F.}\ \bibnamefont {Sauvage}}, \ and\ \bibinfo {author}
  {\bibfnamefont {T.}~\bibnamefont {Fusco}},\ }\href
  {https://arxiv.org/abs/1812.06639} {\enquote {\bibinfo {title} {Overview of
  the european extremely large telescope and its instrument suite},}\ }
  (\bibinfo {year} {2018}),\ \Eprint {http://arxiv.org/abs/1812.06639}
  {arXiv:1812.06639 [astro-ph.IM]} \BibitemShut {NoStop}%
\bibitem [{\citenamefont {Anzuini}\ \emph {et~al.}(2021)\citenamefont
  {Anzuini}, \citenamefont {Bell}, \citenamefont {Busoni}, \citenamefont
  {Motta}, \citenamefont {Robles}, \citenamefont {Thomas},\ and\ \citenamefont
  {Virgato}}]{Anzuini:2021lnv}%
  \BibitemOpen
  \bibfield  {author} {\bibinfo {author} {\bibfnamefont {F.}~\bibnamefont
  {Anzuini}}, \bibinfo {author} {\bibfnamefont {N.~F.}\ \bibnamefont {Bell}},
  \bibinfo {author} {\bibfnamefont {G.}~\bibnamefont {Busoni}}, \bibinfo
  {author} {\bibfnamefont {T.~F.}\ \bibnamefont {Motta}}, \bibinfo {author}
  {\bibfnamefont {S.}~\bibnamefont {Robles}}, \bibinfo {author} {\bibfnamefont
  {A.~W.}\ \bibnamefont {Thomas}}, \ and\ \bibinfo {author} {\bibfnamefont
  {M.}~\bibnamefont {Virgato}},\ }\href {\doibase
  10.1088/1475-7516/2021/11/056} {\bibfield  {journal} {\bibinfo  {journal}
  {JCAP}\ }\textbf {\bibinfo {volume} {11}},\ \bibinfo {pages} {056} (\bibinfo
  {year} {2021})},\ \Eprint {http://arxiv.org/abs/2108.02525} {arXiv:2108.02525
  [hep-ph]} \BibitemShut {NoStop}%
\bibitem [{\citenamefont {Pearson}\ \emph {et~al.}(2018)\citenamefont
  {Pearson}, \citenamefont {Chamel}, \citenamefont {Potekhin}, \citenamefont
  {Fantina}, \citenamefont {Ducoin}, \citenamefont {Dutta},\ and\ \citenamefont
  {Goriely}}]{Pearson:2018tkr}%
  \BibitemOpen
  \bibfield  {author} {\bibinfo {author} {\bibfnamefont {J.~M.}\ \bibnamefont
  {Pearson}}, \bibinfo {author} {\bibfnamefont {N.}~\bibnamefont {Chamel}},
  \bibinfo {author} {\bibfnamefont {A.~Y.}\ \bibnamefont {Potekhin}}, \bibinfo
  {author} {\bibfnamefont {A.~F.}\ \bibnamefont {Fantina}}, \bibinfo {author}
  {\bibfnamefont {C.}~\bibnamefont {Ducoin}}, \bibinfo {author} {\bibfnamefont
  {A.~K.}\ \bibnamefont {Dutta}}, \ and\ \bibinfo {author} {\bibfnamefont
  {S.}~\bibnamefont {Goriely}},\ }\href {\doibase 10.1093/mnras/sty2413}
  {\bibfield  {journal} {\bibinfo  {journal} {Mon. Not. Roy. Astron. Soc.}\
  }\textbf {\bibinfo {volume} {481}},\ \bibinfo {pages} {2994} (\bibinfo {year}
  {2018})},\ \bibinfo {note} {[Erratum: Mon.Not.Roy.Astron.Soc. 486, 768
  (2019)]},\ \Eprint {http://arxiv.org/abs/1903.04981} {arXiv:1903.04981
  [astro-ph.HE]} \BibitemShut {NoStop}%
\bibitem [{\citenamefont {Bell}\ \emph {et~al.}(2020)\citenamefont {Bell},
  \citenamefont {Busoni}, \citenamefont {Robles},\ and\ \citenamefont
  {Virgato}}]{Bell:2020jou}%
  \BibitemOpen
  \bibfield  {author} {\bibinfo {author} {\bibfnamefont {N.~F.}\ \bibnamefont
  {Bell}}, \bibinfo {author} {\bibfnamefont {G.}~\bibnamefont {Busoni}},
  \bibinfo {author} {\bibfnamefont {S.}~\bibnamefont {Robles}}, \ and\ \bibinfo
  {author} {\bibfnamefont {M.}~\bibnamefont {Virgato}},\ }\href {\doibase
  10.1088/1475-7516/2020/09/028} {\bibfield  {journal} {\bibinfo  {journal}
  {JCAP}\ }\textbf {\bibinfo {volume} {09}},\ \bibinfo {pages} {028} (\bibinfo
  {year} {2020})},\ \Eprint {http://arxiv.org/abs/2004.14888} {arXiv:2004.14888
  [hep-ph]} \BibitemShut {NoStop}%
\bibitem [{\citenamefont {Perdrisat}\ \emph {et~al.}(2007)\citenamefont
  {Perdrisat}, \citenamefont {Punjabi},\ and\ \citenamefont
  {Vanderhaeghen}}]{Perdrisat:2006hj}%
  \BibitemOpen
  \bibfield  {author} {\bibinfo {author} {\bibfnamefont {C.~F.}\ \bibnamefont
  {Perdrisat}}, \bibinfo {author} {\bibfnamefont {V.}~\bibnamefont {Punjabi}},
  \ and\ \bibinfo {author} {\bibfnamefont {M.}~\bibnamefont {Vanderhaeghen}},\
  }\href {\doibase 10.1016/j.ppnp.2007.05.001} {\bibfield  {journal} {\bibinfo
  {journal} {Prog. Part. Nucl. Phys.}\ }\textbf {\bibinfo {volume} {59}},\
  \bibinfo {pages} {694} (\bibinfo {year} {2007})},\ \Eprint
  {http://arxiv.org/abs/hep-ph/0612014} {arXiv:hep-ph/0612014} \BibitemShut
  {NoStop}%
\bibitem [{\citenamefont {Sachs}(1962)}]{Sachs:1962zzc}%
  \BibitemOpen
  \bibfield  {author} {\bibinfo {author} {\bibfnamefont {R.~G.}\ \bibnamefont
  {Sachs}},\ }\href {\doibase 10.1103/PhysRev.126.2256} {\bibfield  {journal}
  {\bibinfo  {journal} {Phys. Rev.}\ }\textbf {\bibinfo {volume} {126}},\
  \bibinfo {pages} {2256} (\bibinfo {year} {1962})}\BibitemShut {NoStop}%
\bibitem [{\citenamefont {Kelly}(2004)}]{Kelly:2004hm}%
  \BibitemOpen
  \bibfield  {author} {\bibinfo {author} {\bibfnamefont {J.~J.}\ \bibnamefont
  {Kelly}},\ }\href {\doibase 10.1103/PhysRevC.70.068202} {\bibfield  {journal}
  {\bibinfo  {journal} {Phys. Rev. C}\ }\textbf {\bibinfo {volume} {70}},\
  \bibinfo {pages} {068202} (\bibinfo {year} {2004})}\BibitemShut {NoStop}%
\bibitem [{\citenamefont {Riordan}\ \emph {et~al.}(2010)\citenamefont {Riordan}
  \emph {et~al.}}]{Riordan:2010id}%
  \BibitemOpen
  \bibfield  {author} {\bibinfo {author} {\bibfnamefont {S.}~\bibnamefont
  {Riordan}} \emph {et~al.},\ }\href {\doibase 10.1103/PhysRevLett.105.262302}
  {\bibfield  {journal} {\bibinfo  {journal} {Phys. Rev. Lett.}\ }\textbf
  {\bibinfo {volume} {105}},\ \bibinfo {pages} {262302} (\bibinfo {year}
  {2010})},\ \Eprint {http://arxiv.org/abs/1008.1738} {arXiv:1008.1738
  [nucl-ex]} \BibitemShut {NoStop}%
\bibitem [{\citenamefont {Buckley}\ \emph {et~al.}(2015)\citenamefont
  {Buckley}, \citenamefont {Ferrando}, \citenamefont {Lloyd}, \citenamefont
  {Nordstr\"om}, \citenamefont {Page}, \citenamefont {R\"ufenacht},
  \citenamefont {Sch\"onherr},\ and\ \citenamefont {Watt}}]{Buckley:2014ana}%
  \BibitemOpen
  \bibfield  {author} {\bibinfo {author} {\bibfnamefont {A.}~\bibnamefont
  {Buckley}}, \bibinfo {author} {\bibfnamefont {J.}~\bibnamefont {Ferrando}},
  \bibinfo {author} {\bibfnamefont {S.}~\bibnamefont {Lloyd}}, \bibinfo
  {author} {\bibfnamefont {K.}~\bibnamefont {Nordstr\"om}}, \bibinfo {author}
  {\bibfnamefont {B.}~\bibnamefont {Page}}, \bibinfo {author} {\bibfnamefont
  {M.}~\bibnamefont {R\"ufenacht}}, \bibinfo {author} {\bibfnamefont
  {M.}~\bibnamefont {Sch\"onherr}}, \ and\ \bibinfo {author} {\bibfnamefont
  {G.}~\bibnamefont {Watt}},\ }\href {\doibase 10.1140/epjc/s10052-015-3318-8}
  {\bibfield  {journal} {\bibinfo  {journal} {Eur. Phys. J. C}\ }\textbf
  {\bibinfo {volume} {75}},\ \bibinfo {pages} {132} (\bibinfo {year} {2015})},\
  \Eprint {http://arxiv.org/abs/1412.7420} {arXiv:1412.7420 [hep-ph]}
  \BibitemShut {NoStop}%
\bibitem [{\citenamefont {Kovarik}\ \emph {et~al.}(2016)\citenamefont {Kovarik}
  \emph {et~al.}}]{Kovarik:2015cma}%
  \BibitemOpen
  \bibfield  {author} {\bibinfo {author} {\bibfnamefont {K.}~\bibnamefont
  {Kovarik}} \emph {et~al.},\ }\href {\doibase 10.1103/PhysRevD.93.085037}
  {\bibfield  {journal} {\bibinfo  {journal} {Phys. Rev. D}\ }\textbf {\bibinfo
  {volume} {93}},\ \bibinfo {pages} {085037} (\bibinfo {year} {2016})},\
  \Eprint {http://arxiv.org/abs/1509.00792} {arXiv:1509.00792 [hep-ph]}
  \BibitemShut {NoStop}%
\bibitem [{\citenamefont {Busoni}\ \emph {et~al.}(2017)\citenamefont {Busoni},
  \citenamefont {De~Simone}, \citenamefont {Scott},\ and\ \citenamefont
  {Vincent}}]{Busoni:2017mhe}%
  \BibitemOpen
  \bibfield  {author} {\bibinfo {author} {\bibfnamefont {G.}~\bibnamefont
  {Busoni}}, \bibinfo {author} {\bibfnamefont {A.}~\bibnamefont {De~Simone}},
  \bibinfo {author} {\bibfnamefont {P.}~\bibnamefont {Scott}}, \ and\ \bibinfo
  {author} {\bibfnamefont {A.~C.}\ \bibnamefont {Vincent}},\ }\href {\doibase
  10.1088/1475-7516/2017/10/037} {\bibfield  {journal} {\bibinfo  {journal}
  {JCAP}\ }\textbf {\bibinfo {volume} {10}},\ \bibinfo {pages} {037} (\bibinfo
  {year} {2017})},\ \Eprint {http://arxiv.org/abs/1703.07784} {arXiv:1703.07784
  [hep-ph]} \BibitemShut {NoStop}%
\bibitem [{\citenamefont {Bell}\ \emph {et~al.}(2018)\citenamefont {Bell},
  \citenamefont {Busoni},\ and\ \citenamefont {Robles}}]{Bell:2018pkk}%
  \BibitemOpen
  \bibfield  {author} {\bibinfo {author} {\bibfnamefont {N.~F.}\ \bibnamefont
  {Bell}}, \bibinfo {author} {\bibfnamefont {G.}~\bibnamefont {Busoni}}, \ and\
  \bibinfo {author} {\bibfnamefont {S.}~\bibnamefont {Robles}},\ }\href
  {\doibase 10.1088/1475-7516/2018/09/018} {\bibfield  {journal} {\bibinfo
  {journal} {JCAP}\ }\textbf {\bibinfo {volume} {09}},\ \bibinfo {pages} {018}
  (\bibinfo {year} {2018})},\ \Eprint {http://arxiv.org/abs/1807.02840}
  {arXiv:1807.02840 [hep-ph]} \BibitemShut {NoStop}%
\end{thebibliography}%
\end{document}